# AI-Assisted Conversational Interviewing: Effects on Data Quality and Respondent Experience

Soubhik Barari, Jarret Angbazo, Natalie Wang, Leah M. Christian,
Elizabeth Dean, Zoe Slowinski, Brandon Sepulvado

Methodology & Quantitative Social Sciences
NORC at the University of Chicago

December 1, 2025

**Abstract**

Standardized surveys scale efficiently but sacrifice depth, while conversational interviews improve response quality at the cost of scalability and consistency. This study bridges the gap between these methods by introducing a framework for AI-assisted conversational interviewing. To evaluate this framework, we conducted a web survey experiment where 1,800 participants were randomly assigned to AI 'chatbots' which use large language models (LLMs) to dynamically probe respondents for elaboration and interactively code open-ended responses to fixed questions developed by human researchers. We assessed the AI chatbot's performance in terms of coding accuracy, response quality, and respondent experience. Our findings reveal that AI chatbots perform moderately well in live coding even without survey-specific fine-tuning, despite slightly inflated false positive errors due to respondent acquiescence bias. Open-ended responses were more detailed and informative, but this came at a slight cost to respondent experience. Our findings highlight the feasibility of using AI methods such as chatbots enhanced by LLMs to enhance open-ended data collection in web surveys.

# 1. Introduction

A core tension in survey methodology lies between scalability and adaptability. Standardized, self-administered surveys solve long-standing issues of response incomparability by minimizing interviewer effects and enforcing consistent question wording (Feldman et al., 1951; Fellegi, 1964). However, their rigidity – especially in regards to open-ended questions – limits opportunities for clarification, elaboration, and respondent engagement (Fowler & Mangione, 1990; Schober & Conrad, 1997; Tourangeau et al., 2000). In contrast, conversational interviewing addresses many of these challenges by enabling dynamic clarification, probing, and even real-time validation of researcher interpretations (Schober & Conrad, 1997; West et al., 2018; Hubbard et al., 2020). However, conversational interviewing's dependence on trained interviewers limits scalability and introduces new sources of bias. Moreover, integrating conversational or interactive elements into self-administered web surveys has historically been challenging.

Generative artificial intelligence (AI) is an emerging technology that offers a new opportunity to bridge these divides in web survey methodology. Large language models (LLMs), in particular, with their ability to process and generate human-like text (Bail, 2024), can be integrated into web surveys to enable scalable, self-administered conversational interviewing. Chatbots powered by large language models share the turn-based, dialogic structure of earlier rule-based systems such as ELIZA (Weizenbaum, 1966), but exhibit far more sophisticated reasoning and language capabilities owed to the vast training data used to calibrate the millions, sometimes billions of parameters in their underlying deep neural network models. While similar systems have been referred to by a range of terms in methodological literature – including *AI conversational interviewers*, *AI-powered chatbots*, *conversational AI agents*, *AI interviewers*, and *adaptive interviewers* – we adopt the term *AI chatbots* in this study for clarity and consistency with prior literature. We refer to the broader practice of using these systems to assist (but not fully replace) human interviewers or survey administrators in live data collection as *AI-assisted conversational interviewing*.

A growing body of methodological and applied research demonstrates how AI chatbots can both enhance the respondent experience and, in certain circumstances, improve survey data quality (Xiao et al., 2020; Wuttke et al., 2024; Velez & Liu, 2024). Still, several areas for theoretical development and empirical evaluation remain underdeveloped in this nascent literature, which motivates the present study. First, a clear typology of the capabilities of AI chatbots and how they align with established features of conversational interviewing remains underdeveloped. For instance, chatbots are capable of multiple types of probing, including to elicit depth (e.g., "Can you say more about that?") and to verify the meaning of responses (e.g., "Just to confirm, did you mean…?"). Drawing from the literature on human-led conversational interviewing, we develop and apply a typology of probes to better specify the role AI can play as interview assistant. Second, while much of the integration of chatbots into surveys has centered on probing, LLMs possess a broader range of capabilities. These include live-coding responses (e.g., identifying subjective well-being from open-text entries), as human interviewers must often do for complex, branching surveys in face-to-face data collection. However, there has been little systematic evaluation of how each of these capabilities contributes to data quality. Finally, researchers still lack an understanding of whether AI chatbots perform differently across types of survey content, such as factual versus attitudinal questions.

To answer these questions, we present results from a survey experiment conducted on a conversational AI survey platform. Our design varies both the type of question (factual vs. opinion-based) and the specific chatbot intervention (e.g., with or without probing, and with different types of probes). We evaluate a set of AI-assisted conversational interviewing techniques and their effects on data quality (i.e., the quality of open-ended responses as well as the codes derived from them) and the respondent experience.

In the next section, we situate our study within the literatures on open-ended questions (the question format that our study seeks to improve), web probing and conversational interviewing (which offers tools to enhance open-ended responses, but suffer from limitations), and large language models (which offer a scalable solution to previous limitations). This background then provides the theoretical rationale for our main research questions.

# 2. Background

## 2.1. Open-Ended Survey Questions

Open-ended survey questions enable respondents to answer in their own words rather than using pre-defined response categories (Schuman & Presser, 1979; Tourangeau et al., 2000). The open-ended format eliminates satisficing behaviors induced by closed-ended questions, where the selection, order, and presentation of response categories can influence the respondent's choice, potentially diverting them away from the 'correct' answer (Krosnick & Alwin, 1987). Moreover, open-ended questions can be used to elicit both subjective opinions and factual information: respondents can elaborate on their opinions (e.g., social attitudes) more thoughtfully as well as specify information (e.g., occupation) that may otherwise be burdensome to identify in a long list of choices (Krosnick, 2017).
Open-ended questions have their challenges too. They are more cognitively demanding, requiring more time to formulate and input a response, particularly on mobile devices (Antoun et al., 2017; Couper et al., 2017). Moreover, compared to short close-ended questions, open-ended questions may be more, rather than less burdensome exacerbating respondent satisficing or nonresponse (Krosnick, 1999). Some drawbacks of the open-ended question are specific to self-administered surveys. Without an interviewer to clarify question wording, respondents may provide incoherent or irrelevant answers or avoid answering altogether (Conrad & Schober, 2000; Conrad et al., 2005). Similarly, interviewers cannot clarify responses to ensure accurate interpretation. This creates a quality 'doom loop' that could be avoid through interactive clarification between researchers and respondents.

## 2.2. Conversational Interviewing

Many of the aforementioned issues with open-ended questions arise out of their usage in *standardized interviewing* (SI). While standardized interviewing practices solves the problem of response incomparability due to interviewer effects (Feldman et al., 1951) and question variation (Fellegi, 1964), the rigidity of scripted surveys introduces other threats to construct validity (Fowler & Mangione, 1990). By now, a substantial body of research demonstrates how *conversational interviewing* (CI), also known as *flexible interviewing* (FI), can effectively mitigate such threats.

For instance, Schober & Conrad (1997) demonstrate that allowing both respondents and interviewers to initiate follow-up questions or ‘probes’ reduces comprehension errors and fosters a more consistent understanding of survey questions. Similarly, Suchman & Jordan (1990) and Conrad & Schober (2005) show that conversational techniques enable respondents to elaborate on their answers, improving the accuracy of coding responses into pre-specified categories while also eliciting more detailed and relevant data. Further, West et al. (2018) find that conversational interviewing enhances response quality for income-related questions and other topics without compromising construct validity or introducing significant interviewer effects. More recently, Hubbard et al. (2020) demonstrate that conversational interviewing techniques can be effectively deployed by a variety of professional interviewers. While interviewers with a stronger sensitivity to respondents’ comprehension are more efficient, the technique significantly improves response quality for both opinion and informational questions.

In addition to enabling clarification and elaboration, conversational interviewing creates an opportunity for real-time respondent validation (also called ‘member checking’ in qualitative research practices), where researchers confirm their interpretations or categorizations with respondents (Schober & Conrad, 1997; Birt et al., 2016). This process amplifies the participant’s voice by directly involving them in constructing the researcher’s understanding of responses (Mason, 1997). Unlike traditional methods that defer such validation until after the study concludes, conversational interviewing allows for ‘live coding’ by the researcher and real-verification and correction from the respondent. For instance, the interviewer may first summarize a respondent’s answers after the conclusion of a module or, if a live codebook is being used for a particular question, explicitly state their intended coding, and then request confirmation or feedback from the respondent.

Conversational interviewing is, of course, not without its own drawbacks. The reliance on trained human interviewers may impose substantial costs on researchers (Groves, 2005) and induce effects such as socially desirable responding. Studies suggest that field interviewers, when asked to ‘live code’ information from respondents, are prone to errors like mistyping and mishearing (Olson & Smyth, 2015; West & Blom, 2017). Moreover, integrating conversational or interactive elements into self-administered web surveys has historically been challenging. Conrad et al. (2003) proposed early interactive adaptations into web surveys using hyperlinks and inaction triggers, but these lacked the adaptability and personalization of live interviewing. Similarly, interactive feedback in web forms can improve response accuracy, as demonstrated in Conrad et al. (2005), but it requires a clearly defined concept of accuracy as well as static pre-programmed rules to enforce accuracy (e.g., hours entered into daily timesheet must sum to 24). Research incorporating avatars (or virtual agents) and audio recordings into web surveys also fell short, often disrupting the interview flow, increasing completion time, or failing to enhance the respondent experience (Conrad et al., 2007; Tourangeau et al., 2000).

## 2.3. Web Probing

Traditionally, interactive probing techniques in conversational interviewing have not been possible to fully replicate in the web context, given the requirement to pre-program static questions before fielding online surveys. The closest approximation has been the practice of *web probing*, or inserting a small number of pre-scripted, open-ended follow-ups in an otherwise standardized questionnaire. Originally developed to validate cross-national survey items and enable a form of cognitive pre-testing at scale, web probes typically ask respondents to explain an earlier closed-ended answer. Behr et al. (2017) offer a useful typology of web probes: category-selection probes which invite respondents to justify why they chose a particular response category, comprehension probes which ask respondents to define a term or describe what they believe the question is "really about," and specific probes which focus on a salient detail of the initial answer.

Web probing faces two main constraints. First, because probes are typically attached to closed-ended items, they offer no direct mechanism for improving the content or codability of open-ended answers – the format typically most in need of clarification. Second, their scripted nature precludes real-time adaptation. As Behr et al. (2017) note, web probes lack "the interactivity that would allow spontaneously acting on issues coming up in the probe response or rephrasing a probe that turns out to be problematic."

Still, the empirical literature on web probing yields useful evidence on the effects of follow-up questions in otherwise self-administered surveys. Embedding open-ended probes can slightly increase break-offs, back-tracking, and answer changes to earlier items, though these effects tend to be rare and manageable (Hadler, 2025). The burden appears modest: neither the number of probes nor topic interest markedly harms response quality, completion rates, or respondent satisfaction (Holland & Christian, 2009; Neuert & Lenzner, 2021).

## 2.4. Large Language Models

Conversational interviewing administered by humans, which is tailored to each respondent, is difficult to scale in large-scale web surveys, while previously studied interactive features in web surveys are scalable but lack personalization. Recent methodological and applied social science research suggests that large language models (LLMs) can bridge this gap by providing scalable, personalized, and interactive survey experiences (Bail, 2024).

LLMs, such as GPT-4, are advanced generative AI systems trained on vast amounts of text data. Large parameterization and deep transformer architectures enable LLMs to process and produce coherent, informative, and contextually aware natural language. In turn, LLMs can simulate human-like conversation and reasoning (Vaswani et al., 2023). A novel feature of language models is that they may be directed through the use of instructional prompts (known as *prompt engineering*) rather than programmed code. Researchers can further adapt LLMs for their particular task by simply providing a small curated dataset of inputs and exemplar outputs, a practice known as *few-shot learning* (Brown et al., 2020). This allows them to adapt to new tasks with minimal supervision, often far more efficiently than earlier supervised machine learning models that required large training datasets. In the context of surveys, this parallels how a human interviewer may be trained using instructions, scripts, and example responses (Billiet & Loosveldt, 1988). A key advantage is that LLMs can perform analytical tasks with minimal input from researchers: for instance, tasks like live occupation coding no longer require exhaustive codebooks or large annotated training sets, as was the case with prior supervised approaches (Schierholz et al., 2018; Schierholz & Schonlau, 2021), but can often be accomplished with brief definitions or instructions alone. As such, LLMs offer a scalable and flexible means to support, or even stand in for, human interviewers in conducting dynamic, text-based interviews.

Despite this ability to produce conversational text and demonstrate reasoning abilities, LLMs are prone to a phenomenon known as *hallucination* – that is, misinterpreting instructions or confidently producing incorrect or misleading outputs (Ji et al., 2023). This is a fundamental feature of such generative models that operate with next-token prediction or advanced 'autocompletion' algorithms. AI researchers have cautioned model users not to conflate these capabilities with truth-seeking motivations or genuine comprehension of human inputs. Moreover, it remains essential that practitioners evaluate and monitor such models for systematic hallucinations in their particular applications, rather than assume that a model's strong performance on industry benchmarks will guarantee low hallucination rates in downstream use cases (Bang et al., 2023).

Methodological evaluations of LLMs for open-ended response processing (Mellon et al., 2024) as well as synthetic respondent modelling (Argyle et al., 2023; Bisbee et al., 2024) are expanding, but as von der Heyde et al. (2025) systematically document, there is relatively limited peer-reviewed literature on AI-assisted interviewing itself. That said, some promising methodological results and applications of AI-assisted conversational interviewing have already emerged. Xiao et al. (2020) show that the usage of an AI chatbot can enhance participant engagement. An analysis of over 5,200 free-text responses from a field experiment where half of respondents completed a standard online survey in Qualtrics, while the other half completed a conversational survey administered by an AI-powered chatbot, showed that chatbot-administered surveys elicited significantly more informative, relevant, specific, and clear responses. In another study, Wuttke et al. (2024) compared the quality of responses elicited by a human interviewer versus an AI chatbot in a small-sample, student-based lab study. Respondents rated the chatbot interviews as comparable to human-led ones in overall quality and completeness, with the chatbot condition offering greater efficiency and standardization. However, the authors caution that their findings may not generalize beyond the lab setting, and the presence of student monitors overseeing the interaction limited the ability to assess whether chatbots lower social desirability bias in sensitive topics.

Taken together, current research suggests that AI chatbots can be integrated into web surveys to mirror many of the benefits of human-led conversational interviewing. Yet the field remains in its infancy, and a generalized understanding of when, how, and for whom AI chatbots can meaningfully enhance survey data quality are far from settled. This motivates our current study, which builds on this foundational work to systematically test AI chatbot capabilities in the context of a self-administered web survey.

# 3. AI-Assisted Conversational Interviewing

## 3.1. Approaches

Building on the previously mentioned literatures, we introduce and operationalize two broad approaches for leveraging AI technologies such as large language models in web survey data collection: live coding and probing.

*AI-assisted live coding* refers to an AI chatbot's ability to detect concepts within open-ended answers in real time, as a field interviewer might (West & Blom, 2017) without the risks of human error (Olson & Smyth, 2015). This functionality draws on machine learning techniques such as text classification, sentiment analysis, and topic modelling (Puri & Catanzaro, 2019). Unlike post-hoc automated coding methods using supervised learning or language models (He & Schonlau, 2022; Gweon & Schonlau, 2024), live coding occurs during data collection, enabling immediate usage for survey branching respondent validation. To facilitate live coding, a standard codebook can be provided to the chatbot with examples (few-shot learning), with only definitions (zero-shot learning), or omitted entirely to prompt on-the-fly category discovery (unsupervised learning).

*AI-assisted probing* refers to an AI chatbot's ability to generate follow-up questions based on a respondent's answer to a seed question, mirroring practices in human interviewing (Billiet & Loosveldt, 1988; Groves et al., 2009). In typical self-administered surveys, all probes must be hard-coded using static logic. In contrast, AI chatbots can be trained to decide when and how to probe based on prior responses or survey paradata.

In this study, we further operationalize three types[1] of AI-assisted probes inspired by earlier typologies of conversational interviewing (Suchman & Jordan, 1990) and web probing (Behr et al., 2017):

1. *Confirmation probes* allow the respondent to confirm whether the chatbot's live coding aligns with the respondent's intended meaning*,* A confirmation probe can be asked as a yes/no question where the chatbot categorizes an answer and asks for confirmation. If multiple categories are detected or if there is uncertainty, the chatbot could ask the respondent to select from a list of possible categories.

2. *Elaboration probes* invite the respondent to provide more detail about an open-ended answer they have given, elaborate on their reasoning, or offer more specific information.

3. *Relevance probes* are used to improve the relevance or interpretability of the original response provided.

We acknowledge that it is possible to administer hybrid probes that overlap between multiple categories. In some cases, one type of probe may accomplish the goal of another, as it might be necessary to request elaboration to determine relevance, or vice versa. For example, in order to clarify whether 'space' refers to 'green space' (relevant) or 'outer space' (irrelevant), a probe may request both elaboration and relevance.

## 3.2. Research Questions

Our study evaluates the effectiveness of AI-assisted conversational interviewing, as operationalized through the methods of live coding and probing. We investigate three core questions:

**RQ1.** How accurately can AI chatbots live-code open-ended survey responses?

[1] See Table A6 in the supplementary appendix for examples of different probes.

Although LLMs have been evaluated for post-hoc classification (Mellon et al., 2024; Heyde et al., 2025) and supervised machine learning algorithms for live classification (e.g., Schierholz et al., 2018; Schierholz & Schonlau, 2021), few studies have evaluated LLMs for live classification. Our study aims to establish a performance floor – that is, how well a chatbot with minimal prompting and no comprehensive codebook can produce accurate classifications.

While existing studies do not replicate our exact setup, they offer useful benchmarks to help calibrate expectations. Schierholz et al. (2018) reported agreement rates between supervised learning algorithms and human coders on occupation codes that ranged from 72.7% to 93.1%, depending on question format and dataset. By comparison, recent evaluations of LLMs have yielded 91-94% classification accuracies on open-ended survey responses, including German-language responses about survey motivation and English-language "most important issue" questions, with slightly lower but still promising results in zero- and few-shot prompting settings. In the context of real-time classification, Schierholz and Schonlau (2021) found that 72.4% of respondents successfully selected an occupation provided by a supervised machine learning algorithm's during the interview. In our case, we might then expect high, but variable accuracy in our real-time context which spans multiple question types.

**RQ2.** Can AI chatbots improve the quality of open-ended responses through probing?

Prior evidence from conversational interviewing (West et al., 2018; Suchman & Jordan, 1990), web probing (Behr et al., 2017) and early AI chatbot evaluations (Wuttke et al., 2024) suggests that dynamic, tailored probes will increase the informational content in open-text responses. Accordingly, we hypothesize a positive effect of probing on open-ended response quality.

**RQ3.** Do AI chatbots affect the overall respondent experience?

While web probing studies suggest limited effects on satisfaction, breakoff, or burden (Neuert & Lenzner, 2021; Hadler, 2025), AI chatbots may introduce additional delay, contribute to response latency, or increase cognitive load relative to a static web survey experience. We treat this as an open empirical question.

The next section details our experimental design to formally evaluate the AI-assisted interviewing approach along these dimensions.

# 4. Methodology

## 4.1. Study Design

In order to evaluate the impacts of AI-based coding and probing on data quality, we designed and administered a survey on a chat-based interface with experiments embedded within each of four distinct question groups. Table 1 summarizes our design, including the flow of the four question modules and their variations in format across experimental conditions. Reference questions were drawn from major surveys as were codebooks to develop the coding frame for live coding and to inform the close-ended response categories.

## Table 1. Overview of Experimental Design

| Survey Module | Seed Question Format Across Conditions<br>Control: No Probes | Treatment 1: Confirmation Probes | Treatment 2: Elaboration Relevance Probes | Coding Dimensions | Reference |
|---|---|---|---|---|---|
| (Exp 1) Most Important Issue | Open-Ended | Open-Ended | Open-Ended | Political issues | Gallup Tracking Poll |
| (Exp 2) Economic Conditions | (i) Close-Ended (Sentiment)<br>(ii) Open-Ended (Reason) | (i) Open-Ended (Sentiment)<br>(ii) Open-Ended (Reason) | Open-Ended | (i) Sentiment (Pos/Neg)<br>(ii) Reason | Pew American Trends Panel |
| (Exp 3) Preferred News Source | Close-Ended | Open-Ended | Open-Ended | News Outlets | NORC AmeriSpeak Profile Survey |
| Demographics | Close-Ended (Age, Gender, Education, Employment) | | | N/A | N/A |
| (Exp 4) Main Occupation (Among Employed) | Close-Ended | Open-Ended | Open-Ended | Occupation Codes | BLS Standard Occupational Classification |
| Respondent Experience | Close-Ended (Quality, Ease, Satisfaction, Frustration) | | | N/A | N/A |

Each of the within-survey experiments included 1-2 open-ended questions on commonly surveyed topics: (1) most important national issue, (2) economic evaluation, (3) news sources, and (4) occupation. A complete questionnaire can be found in Appendix Table A1.

This design, though complex, allowed us to better generalize conversational AI's effects on data quality across different contexts. First, the formats of the control condition question reflect different real-world survey practices: short ordinal lists (economic sentiment), long nominal lists (23 standardized occupational codes), choices with varying granularity (news sources which may include apps or outlets), and open-ended questions requiring ex-post coding (national issue). Second, the four experiments spanned factual responses (news source, occupation) and subjective opinions (national issues, economic evaluation). Third, conceptual categories varied, covering topics (national issues), sentiment (economic sentiment), proper nouns (news source), and classifications (occupation), highlighting the diverse challenges for chatbot coding.

The baseline control question in each experiment was standardized to all respondents with no probing or live coding, while the two treatment versions each administered a specific type of probe based on the initial (seed) question and response. In Treatment 1, the chatbot could only deploy confirmation probes while in Treatment 2, the chatbot was configured to deploy only elaboration or relevance probes.[2] Figure 1 illustrates the exact user interface seen by respondents in different standardized and conversational conditions along with real probes of each type administered in the conversational interviewing conditions.[3]

**Figure 1. Examples of AI-Assisted Probes in Research Design**

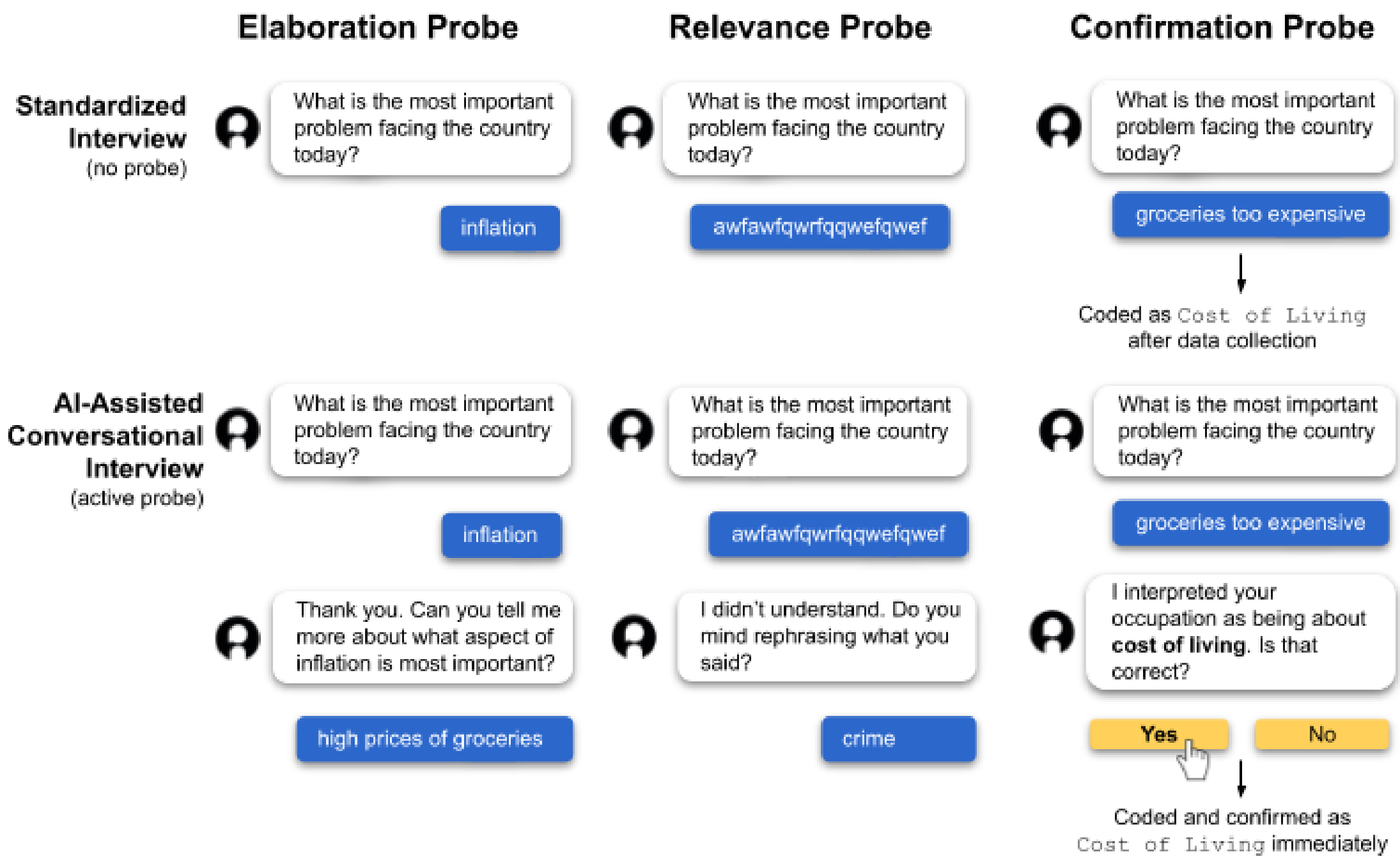

In the second treatment condition, confirmation probes were configured to activate upon the coding of a category for the particular concept of interest (e.g., issue) explicitly sourced from a reference question and/or codebook (e.g., major issue categories identified in the national Gallup tracking poll).[4] Noting in earlier literature that too many probes may increase overall nonresponse (Behr et al., 2012), guardrails were placed in the chat interface such that the chatbot could administer at most one probe per each seed question in any treatment condition.[5]

Finally, to explore heterogeneity in response quality effects, we collected demographic information and asked about the survey experience through a series of Likert-scale questions at the end of the survey. Details and results for heterogeneous treatment effects are given in Appendix Section A4.

---

[2] Ideally, our experiment would unbundle relevance and elaboration probing into two separate conditions, but we combined them into a single treatment due to statistical power concerns and limitations in the user interface with our selected conversational AI platform.
[3] Appendix Figure A1 illustrates how the large language model was configured in order to perform different probes, incorporating content from the seed question and response.
[4] If multiple categories were coded, the chatbot sampled one for confirmation probing. Examining the chatbot metadata, we found that in most responses for most questions, only one category was coded.
[5] Due to platform limitations, indicators for the type of probe administered were not available. Hence, after data collection concluded, coders re-classified the type of each probe in Treatment 2.

## 4.2. Conversational AI Platform

We fielded our experiment on the conversational AI platform Inca, with an interface structurally identical to that depicted in Figure 1. The underlying LLM in our experiment is SmartProbe (Seltzer et al., 2023), a model fine-tuned from the InstructGPT family (Ouyang et al., 2022) which is a collection of models themselves fine-tuned from the GPT-3 foundational model (Brown et al., 2020). Further details on the structure of the SmartProbe model, including known information about its model settings, fine-tuning procedures, and evaluation metrics can be found in Appendix Section A1.

Importantly, though we provided the name of each coding category in our question-level prompts to the chatbot, we did not further provide examples of responses belonging to each category – a form of *zero-shot learning* (Kojima et al., 2022) – for the purposes of a conservative 'off-the-shelf' measure of performance.

To monitor and mitigate any risks of respondents encountering *hallucinations*, or irrelevant content generated by the LLM due to misinterpretations of probing instructions or input responses, we tested our chatbot extensively prior to fielding with sample inputs. We did not encounter any instances of hallucination in the generated probes. Similarly, when reviewing probes generated during the actual study, we only found 2 instances of off-topic probes suggesting hallucinations did not occur at a scale to systematically impact either data quality or respondent experience.

## 4.3. Fielding

Respondents in our study ($n$ = 1,800) were recruited into the Inca platform from a proprietary non-probability panel developed by Prodege, an online market research company, and quota-sampled to match marginal gender, age, and education totals of the U.S. adult population according to the 2020 American Community Survey (ACS). Participants accessed our conversational AI platform and, after providing consent (including acknowledgment of potential interaction with an AI agent), were randomly assigned to one of three experimental conditions.

The survey was fielded on July 15th, 2024 and concluded on July 17th, 2024. Fielding concluded when 601 complete interviews (chats) were collected across each condition, resulting in 1,803 complete interviews, 195 partial interviews (attrition during the survey), and a total of $n$ = 1,998 interviews altogether. 65% of complete interviews were conducted on a smartphone device, 32% were completed on a Desktop device, and the remaining 3% were completed on other devices such as tablets. Details on the composition of respondents as well as summary statistics of each outcome measure can be found in the Appendix.

## 4.4. Outcomes

**Coding Performance (Respondent Confirmation).** We first evaluate the confirmation probing (in Treatment 2) in terms of *accuracy* with respect to the respondent's own confirmation of the coded category – codings met with a "no" response or undetected categories requiring a categorical confirmation were considered inaccurate. Additionally, we measured *precision* or the degree of false positive error (categories detected but not confirmed) and *recall* or the degree of false negative error (categories not detected but later confirmed by respondents).

**Coding Performance (Human Coder Agreement).** We also compared the chatbot's live coding to a team of three human coders' labels created after data collection, according to the same codebook, on a sample of 100 responses for each question. Discrepancies could occur, for instance, due to acquiescence bias (Krosnick, 1999), where respondents are biased towards confirmation even if the coded category is not accurate.

**Response Quality (Qualitative).** We measure response quality in each open-ended response across our experiments based on both qualitative human assessments of response quality and quantitative measures of textual information.

Qualitive indicators of response quality were defined and assessed by a team of three human coders. First, the team of coders inductively constructed definitions of 'response quality' using independent samples of 100 open-ended responses divided evenly across treatment conditions from each question group. Definitions were refined until agreement could be reached on the exact meaning of each and whether they could be reliably identified. The final criteria, similar to those developed in closely related studies (Xiao et al., 2020; Wuttke et al., 2024), are as follows:

- *Relevance.* Respondent provides a response that answers the question (e.g., stays on topic, does not refer to other questions).
- *Specificity.* Respondent provides specificity in their answer (e.g., providing examples or proper nouns, if applicable, rather than referring to abstract concepts).
- *Explanation.* Respondent provides explanations, motivations, or reasons for their response (only applicable for opinion questions, i.e. 'Most Important Issue' and 'Economic Conditions').
- *Completeness.* Respondent provides a response that fully answers the question and does not omit any part of the question, if there are multiple parts.
- *Comprehensibility.* Respondent provides a response largely free of major spelling, grammatical, or other syntactic errors that deter human understanding of the response.
- *Concision.* Respondent does not provide duplicate or redundant information.

These criteria formed a codebook that was then applied to a larger sample of open-ended responses from each question-level experiment among both Control and Treatment 1 respondents. Crucially, the coders applied these criteria to the combined text of the seed/probe responses in Treatment 1 and were blind to the treatment status of each respondent that they coded. A second round of coders repeated the process on 100 pre-probing responses in Treatment 1. This allowed us to compare response quality before and after probing and to assess whether the chatbot targeted low-quality responses or probed indiscriminately.

**Response Quality (Quantitative).** As a supplement to human-coded quality criterion, we created a slate of quantitative measures to characterize the total informational content in each response. First, we measured the *lexical diversity* operationalized as the number of unique words in each response as a ratio of the total number of words. A higher score indicates a more diverse vocabulary which may signal that the respondent used a wide range of words, potentially providing more detailed and varied information. Second, we measured the *Shannon entropy* of responses, defined as the uncertainty or unpredictability of word occurrences within each response. This metric is calculated as:

$$H(x) = -\sum_{i=1}^{n} p(x_i) \log (p(x_i)),$$

where *n* is the number of unique words in the given response text represented by $x$ and $p(x_i)$ denotes the probability of the *i*th word $x_i$ appearing in the response (operationalized as the number of occurrences over the total word count in the response). Greater Shannon entropy estimates mean that the response contains a wider range of words with more even distributions, implying a richer and more informative response. Conversely, lower entropy estimates indicate a less varied use of words and uneven values of $p(x_i)$, suggesting that the response may be more repetitive or less informative.

We also measured the *Kullback-Leibler (KL) divergence*, which quantifies how the word distribution in a respondent's response diverges from the overall word distribution across all responses. This is operationalized as:

$$KL(x) = -\sum_{i=1}^{n} \log \frac{p(x_i)}{q(x_i)},$$

where $q(x_i)$ denotes the probability of word $x_i$ appearing in any respondent's response for that particular question (operationalized as the total count of that word in the sample over the total word count for that question across all respondents). In contrast to Shannon entropy which provides a conservative absolute measure of informativeness, a greater KL divergence means the respondent's word usage is more unique relative to other responses in the overall sample. A lower KL divergence suggests that the response closely follows the typical word distribution, which might indicate less uniqueness or specificity in the information provided.

Finally, we capture the *total number of words* and *unique words* in each response. If a participant responds at all to a probe, we should obviously expect a higher total word count in the combined seed and post-probe response string. However, more words and varied word choice in the post-probe response alone may indicate that the probe elicited more information from the participant.

**Respondent Experience (Behavioral).** We measured the quality of respondents' survey experience both behaviorally and attitudinally.

Our behavioral measure of experience was attrition (or, interchangeably, dropout) during or after each question-level experiment. Attrition is considered a strong barometer of respondent experience. If a respondent drops out, it suggests that their experience was poor enough to deter them from continuing the survey (Groves et al., 2009). The advantage of this measure is that it allows us to isolate the effect of each question on respondent experience. However, it is important to note that question order may also affect dropout rates, as earlier questions could more strongly influence a respondent's decision to leave the survey. Due to limitations of the platform, we could not measure completion time for individual questions, though we were able to observe the total duration of each interview. While response times can be interpreted a proxy for respondent inattention and other sources of measurement error, in practice, it can be difficult to disentangle from natural variations in survey-taking speed not related to engagement (Yan & Olson, 2013). In our context, response times are a function of both respondent's time spent answering and the chatbot's latency in generating probes. Nonetheless, we provide a summary of overall survey timing in Appendix Table A3 and Table A4.

**Respondent Experience (Attitudinal).** Attitudinal reports of respondent experience were measured using a series of 5-point Likert scale questions at the end of the survey along different dimensions: *Quality, Ease, Frustration,* and *Satisfaction* (full question wording in Appendix). Though these attitudinal measures provide granular evaluations of the respondent's experience, there are limitations. Since they are self-reported at the end of the survey, they may be biased upwards, as respondents who had a lower evaluation of the survey experience may have already dropped out. Additionally, responses to these questions at the end of the survey may be subject to satisficing, where respondents provide satisfactory rather than optimal answers due to fatigue. Despite these limitations, consideration of both behavioral and attitudinal measures allows for a more robust understanding of how respondents experience a survey (Groves et al., 2009).

### 4.5. Analysis

All analyses were conducted using the R programming language. We estimated treatment effects on the aforementioned outcomes using OLS regression models, both without and with control covariates obtained from Prodege about our panelists: work status (except for analyses related to the occupation question), gender, household income, educational attainment, and device type. Estimates are presented with confidence intervals adjusted for multiple comparisons using the BHq correction (Benjamini & Hochberg, 1995), with the family set at the level of each figure.[6]

## 5. Results

### 5.1. Live Coding and Confirmation Probing (RQ1)

We begin by examining the results related to confirmation probing in Treatment 1. Table 2 and Table 3 evaluate live coding in two ways: according to the respondent's own confirmation and according to independent human coders' aggregated coding of each relevant concept.

**Table 2. Coding Performance (Respondent Confirmation) in Confirmation Probing Condition (Treatment 1)**

| Performance Metric | Most Imp. Issue | Econ. Cond. (Sentiment) | Econ. Cond. (Reason) | Pref. News | Main Occu. |
|---|---|---|---|---|---|
| Accuracy | 73.5% | 96.2% | 80.9% | 66.1% | 84.8% |
| Precision | 95.9% | 96.2% | 96.2% | 93.0% | 91.8% |
| Recall | 73.6% | 92.2% | 80.7% | 61.2% | 85.1% |

[6] For the purposes of interpretability, we present estimates produced by linear regression even when outcomes are binary (e.g., response quality indicators), per the recommendations of Gomila (2021). All such results are substantively similar when replicated using average predictive comparisons estimated from logistic regression.

Accuracy refers to the % of responses where respondent agreed with AI-coded category or non-coding; precision refers to % of AI codings with response 'yes' in confirmation probe; recall refers to the % confirmed category incidences that were coded by the AI.

With the exception of news sources, where accuracy and recall both fell below 70%, the chatbots performed better than random guesses (>70% correct) without additional training, based on respondents' confirmations. The chatbot's recall for the preferred news source question was low, reflecting a greater need for respondents to manually confirm a category that was missed by the chatbot (path (b) under Treatment 2 in Figure 1). Nevertheless, the chatbot's initial selection of response categories in Treatment 2 correlates strongly with those directly selected by Control respondents in close-ended questions (Figure A9).

Respondents tend to agree more often with the chatbot's coding than do independent human coders. For instance, 81% of respondents confirm that the chatbot's characterization of their economic sentiment reasons is correct, while only 72% of the human codings align with the chatbot's coding. What accounts for this discrepancy? As shown in Table 3 (second row), the precision of coder assessments is, on average, 20% lower than that of respondents' confirmations (second row in Table 2). In other words, respondents tend to favor the "yes" choice in a confirmation probe, a sign of acquiescence bias and a common issue in surveys.

### Table 3. Coding Performance (Human Coder Agreement) in Confirmation Probing Condition (Treatment 1)

| Performance Metric | Most Imp. Issue | Econ. Cond. (Sentiment) | Econ. Cond. (Reason) | Pref. News | Main Occu. |
|---|---|---|---|---|---|
| Accuracy | 73.1% | 90.9% | 72.0% | 71.0% | 82.6% |
| Precision | 79.7% | 90.9% | 66.7% | 60.9% | 77.9% |
| Recall | 72.0% | 90.9% | 69.6% | 56.0% | 76.8% |

Accuracy refers to the % of responses where majority human coding matched coded category / non-coding; precision refers to % of AI codings where majority human coding agreed with category; recall refers to the % of human-coded category incidences that were coded by the AI.

Supplementary analyses (Appendix Figure A6) provide further evidence of acquiescence bias: while overall category rates are positively correlated between coders and respondents ($\rho$ = 0.48–0.87), "None of the above" is a consistent outlier, with respondents selecting it under 5% of the time despite coders applying it 10–40% of the time.

What are the consequences of using live coding and confirmation probing relative to simply asking close-ended questions? In supplementary analyses (Figure A7 – Figure A11) comparisons of response categories in other conditions to how they're either confirmed by respondents or coded by coders in Treatment 2 reveal strong correlations in category incidence ($\rho$ = 0.75–0.88 across questions). This suggests that the usage of confirmation probing may not necessarily induce severe construct validity errors as sometimes occurs through probing (Kuha et al., 2018). Moreover, when respondents do not confirm the chatbot's suggested response, they often select a thematically similar category (e.g., 'Inflation' rather than 'Employment/Jobs'), suggesting the chatbot is unlikely to grossly misrepresent a respondent's intended meaning (Appendix Figure A12).

Finally, coding performance remained consistent across device types, with economic tone and occupation showing the highest levels of accuracy (see Table A5). However, desktop users tended to affirm the chatbot's coding more frequently for economic reasons. It is important to note that these differences do not necessarily indicate greater acquiescence bias on one device type versus another. The original content of the responses – and consequently the chatbot's coding accuracy – may differ between devices, which could also influence respondents' confirmation behavior. Moreover, survey fatigue, and correspondingly the willingness to acquiesce particularly on later questions, may operate differently across device type.

## 5.2. Rates of Elaboration and Relevance Probing (RQ2)

Before presenting the effects of probing on response quality, we first examine patterns of when probes were and were not triggered in Treatment 2. As Table 4 shows, for only a tiny minority (1-4%) of seed responses to each question were probes not administered. Moreover, the vast majority of probes (up to 98% for the occupation question) could be strictly characterized as elaboration probes, rather than relevance probes. This should be entirely expected, since the baseline relevance of seed responses nearly hits the ceiling, almost 100% for the most important issue and economic conditions questions (baselines by condition shown in Appendix Figure A3), whereas explanation and specificity tends to be present slightly less often.

Frequent triggering of elaboration probes suggests that the chatbot may be perceptive of the specific quality 'needs' in the respondent's seed response, tailoring the type of probe to the dimension of quality in deficit. A comparison of response quality between seed responses and post-probe responses in Appendix Figure A13 shows this is true for some questions – with the most important issue question, for instance, the average seed response that does receive a probe is one of lower quality across the six criteria than the average seed response that induced a probe. This pattern, however, does not extend to the news question where no-probe seed responses have significantly lower completeness and relevance than probed seed responses, though this can only be garnered from 6 seed responses who did not receive a probe.

Finally, as we noted, our coders discovered that there were instances where the chatbot delivered "hybrid probes" with elements of both elaboration and relevance, particularly for the news question (15 such instances).

### Table 4. Rates of Probes Triggered in Treatment 2 (Elaboration/Relevance Probing)

| Probe Triggered | % (n) of Seed Responses Where Probe Was Triggered | | | |
|---|---|---|---|---|
| | **Most Imp. Issue** | **Econ. Cond.** | **Pref. News** | **Main Occu.** |
| Elaboration Probe | 94% (622) | 85% (538) | 68% (414) | 98% (361) |
| Relevance Probe | 2% (13) | 11% (69) | 29% (175) | 1% (5) |
| Hybrid Probe | 0% (1) | 1% (6) | 2% (15) | 0% (0) |
| No Probe | 4% (24) | 3% (17) | 1% (6) | 0% (1) |
| Probe Error | 0% (0) | 0% (0) | 0% (2) | 0% (0) |

Counts exclude respondents who dropped out prior to question being asked (and thus were given no opportunity for a probe). Examples of specific probes can be found in Appendix Table A6.

## 5.3. Effects of Elaboration and Relevance Probing on Response Quality (RQ2)

Next, we turn to the effects of elaboration/relevance probing on the overall quality of open-ended responses (combining the seed and post-probe response) for each question.

Figure 2 shows that there is, on average, a substantial increase in rates of specificity and explanation criteria of the most important issue and economic evaluation reasoning open-ended responses when respondents are exposed to probing. Probing introduces the risk of redundancy between the seed and post-probe response, but concision does not, on average, decrease when probes are delivered in Treatment 2. Between the Control and Treatment 2 conditions, no other response quality criteria experienced a difference in either direction. The results from

Figure 2 are consistent with results from a pre-post design – comparing coded rates of response quality in the seed response to the post-probe response within individual (Appendix Figure A14).

**Figure 2. Effects of Elaboration/Relevance Probing (Treatment 2 vs. Control) on Qualitative Response Quality Measures**

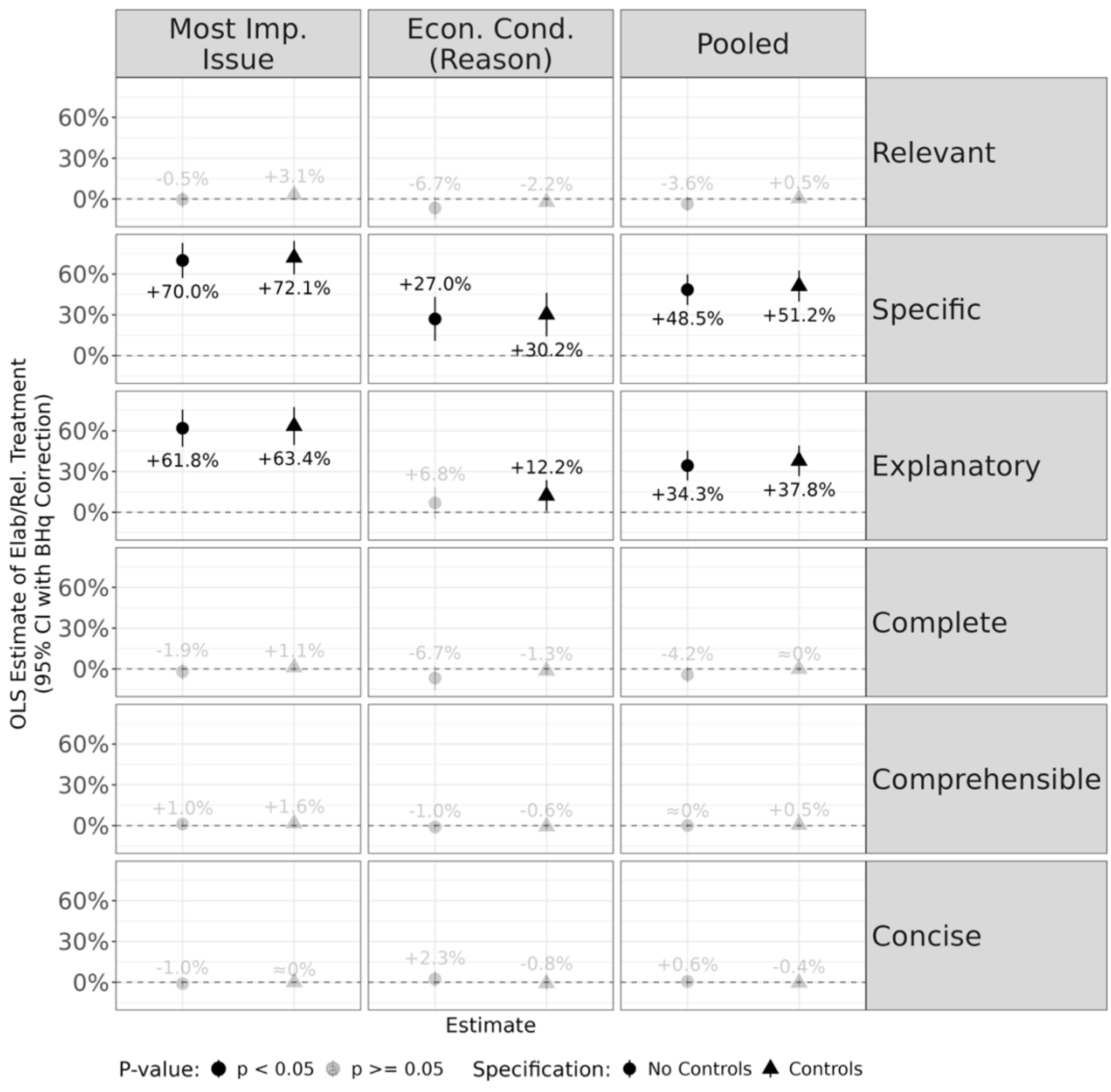

Figure 3 presents the impacts of receiving probes in Treatment 2 on information content, operationalizing the outcome to be either the post-probe or combined seed/post-probe response, with seed responses in the Control condition as the baseline. It is trivial that both total and unique words increased with probing when the outcome is the combined response, though the mean word count in the post-probe response was higher than the mean word count in the seed response. While probing increased the Shannon Entropy in the subsequent response, it did not increase (and sometimes decreased) the KL divergence, meaning that new words in the post-probe response did not significantly deviate from the overall distribution of words, or 'contribute new information' to the sample. Lexical diversity also did not experience an effect, suggesting that new concepts may not have been systematically introduced post-probe that were not referenced in the first response. Subgroup-level estimates of treatments effects exhibit nearly no variation across these measures.

**Figure 3. Effects of Elaboration/Relevance Probing (Treatment 2 vs. Control) on Quantitative Response Quality Measures**

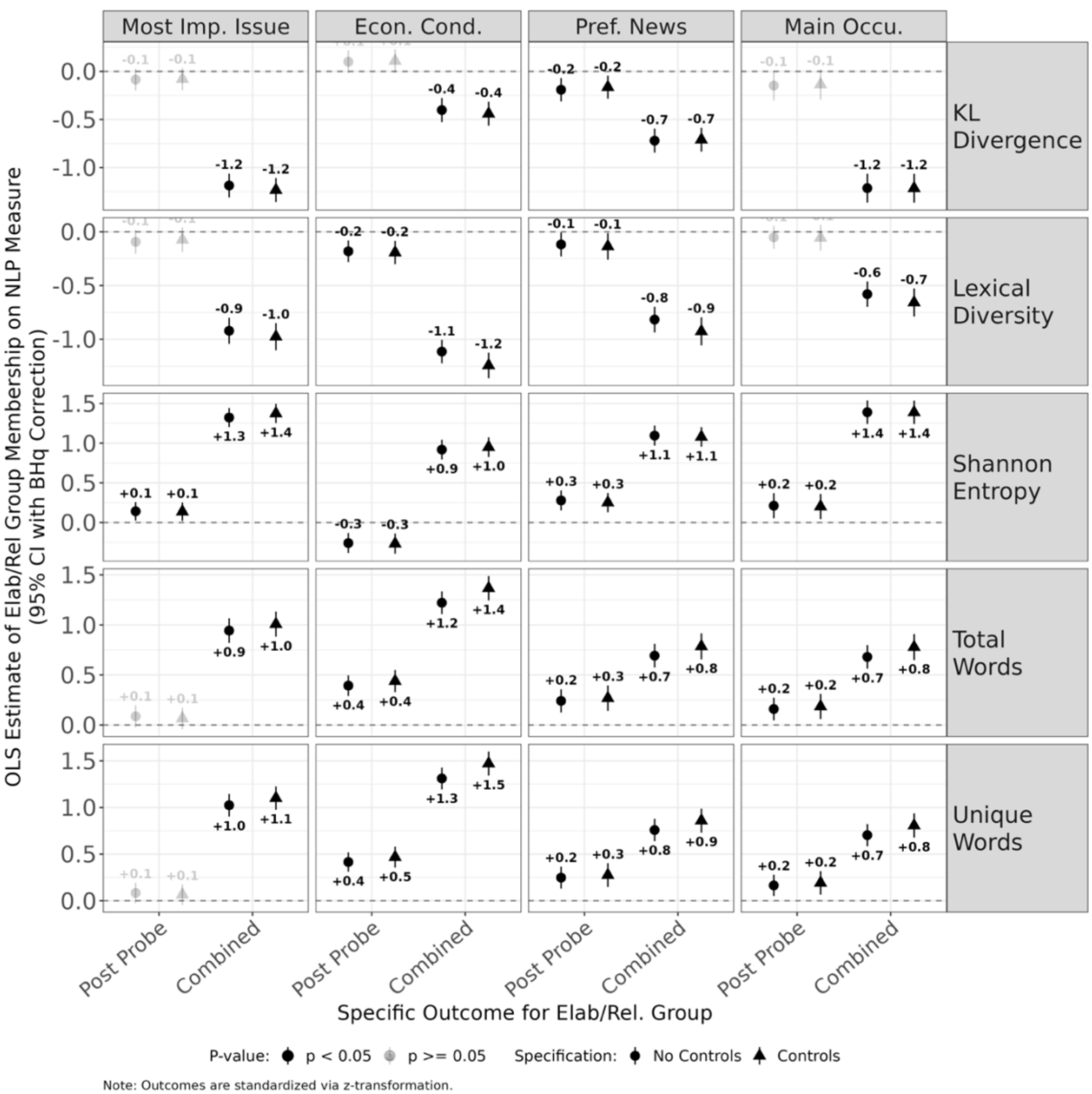

## 5.4. Effects of Probing on Respondent Experience (RQ3)

Completion patterns suggest probing does impact respondent experience (Appendix Table A2): dropouts were approximately twice as high in Treatment 1 (Confirmation Probing) compared to the Control Condition ($n$ = 88 vs. $n$ = 43). Similarly, interview duration was longer for both treatment conditions, particularly in Treatment 2 (Elaboration or Relevance Probing) where the average chat lasted twice as long (6 min.) as in the Control Condition (3 min.) (Appendix Table A3 and Table A4).

Figure 4 provides a comparison of attrition (operationalized either as respondent drop-out immediately after or any time after exposure to a question) across questions in both Treatment 1 and Treatment 2 relative to rates in the Control condition. Confirmation probing does not appear to increase attrition propensity any time during the survey, while elaboration or relevance probing only increases the likelihood to drop out by 2-3% for the first question. That subsequent probes do not induce further attrition suggests that either respondents habituate to the probing design or that the first question probes weeds out respondents who cannot habituate, or both (Behr et al., 2014). We did not detect any heterogeneity in attrition effects across any subgroup for any question or treatment condition. Despite the greater attrition rates induced by elaboration/relevance probing, coded response categories between Treatment 1 and Treatment 2 remain highly correlated across questions (Appendix Figure A9).

**Figure 4. Effects of Probing (Treatment 1 or 2) on Respondent Attrition**

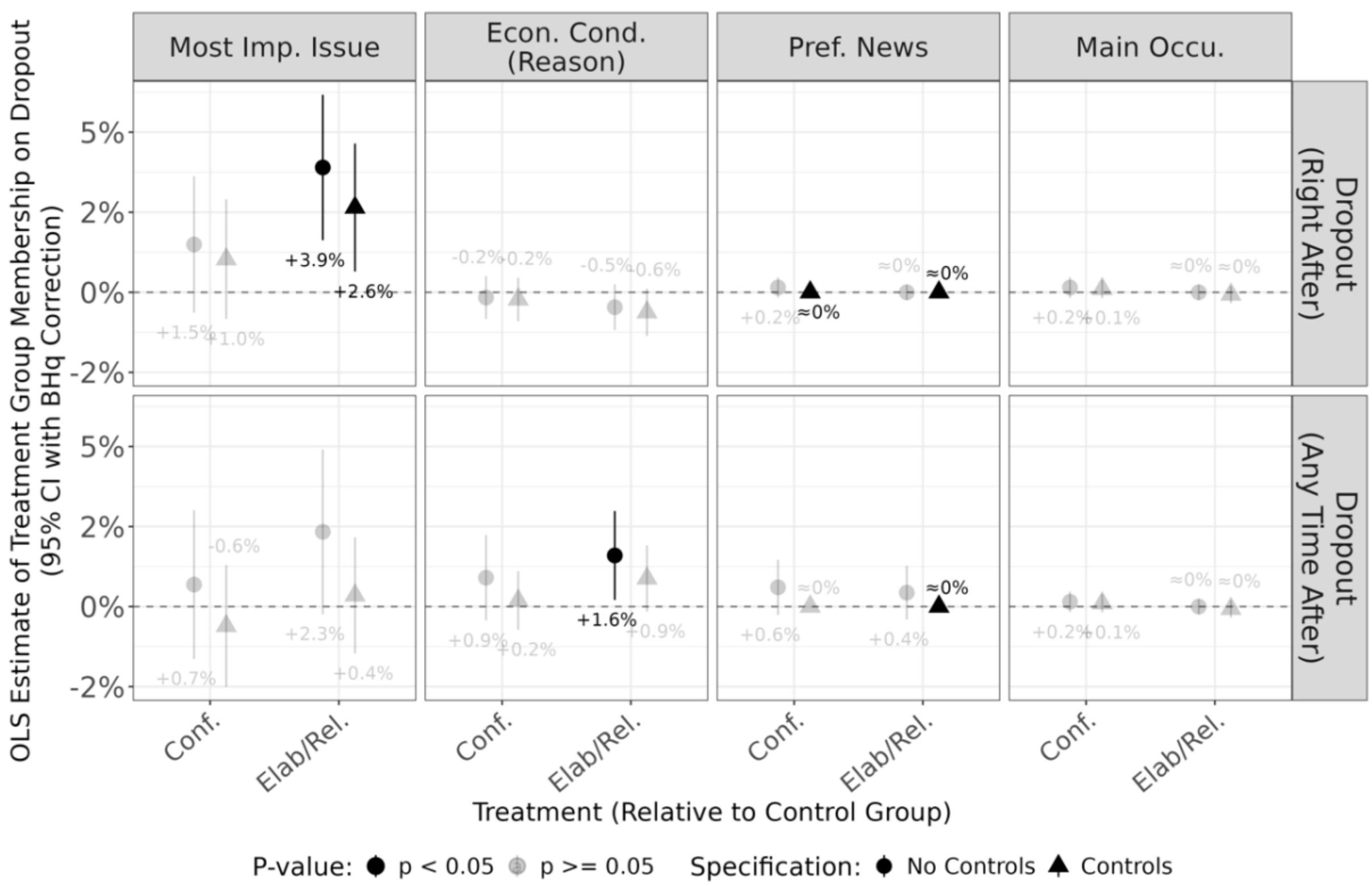

Note: Higher levels correspond to higher rates of attrition. Control covariates include work status, gender, education, age, and device type.

Lastly, we present how exposure to either treatment bundle of probes shapes the overall self-reported survey experience. As Figure 5 shows, we find statistically significant but substantively small negative effects (in all cases, a movement of less than 0.05 on a normalized 0-1 scale) of receiving the full dosage of elaboration/relevance probing (Treatment 2) on ease, frustration, and satisfaction. Relative to the Control interview, the full sequence of confirmation probes introduced in Treatment 1 has no effect on any self-reported dimensions of respondents' experience. These results are almost certainly biased in a positive direction: respondents who 'survive' to evaluate the survey experience are likely to have a less negative experience than respondents who did not complete the interview.

**Figure 5. Effects of Probing (Treatment 1 or 2) on Self-Reported Respondent Experience**

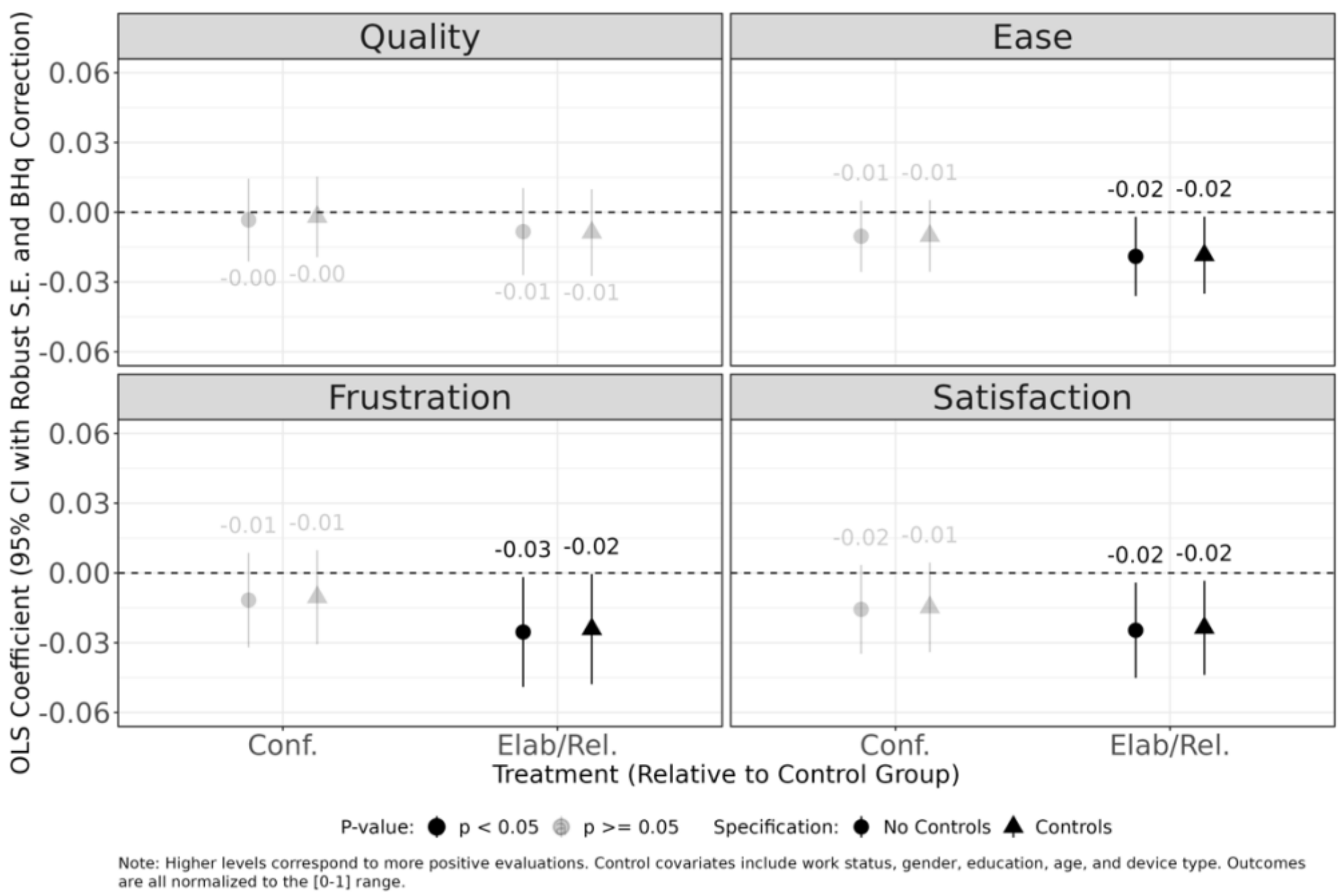

On respondent experience, we do observe differences in treatment effects by demographic group. In particular, when exposed to elaboration/relevance probing, mobile respondents tend to report lower ease of use and frustration, while desktop respondents only rate lower frustration. Older respondents (aged 60+) are the only age group to significantly rate lower ease of use as are respondents with a Bachelor's degree. While such subgroup-level negative effects are statistically significant, as with sample-wide effects, they remain small in overall magnitude.

# 6. Discussion

We have conducted one of the first evaluations of AI-assisted conversational interviewing on survey data quality and respondent experience. We begin by summarizing our findings and considering their practical implications for survey researchers.

## 6.1. Findings and Practical Implications

Our results across all three research questions indicate mixed effects: using 'off-the-shelf' AI-assisted conversational interviewing can provide some modest data quality benefits at a minor cost to the survey experience.

For live coding open-ended responses (RQ1), we find evidence that LLMs can perform with a high degree of accuracy, precision, and recall across both opinion and factual reporting questions. Our chatbot achieved high accuracy in coding binary economic sentiment and occupation, but struggled more with news sources, possibly due to variability in naming conventions (e.g., “NYT” vs. “The New York Times”) and acronym ambiguity (e.g., “CNN”, an abbreviation for both a news network and type of machine learning model). Overall, AI-assisted coding performed better than chance, with a 74% confirmation rate among desktop users, slightly higher than 72.4% yielded by Schierholz et al. (2018)’s trained supervised learning algorithms. The errors we have identified in our evaluation could concretely inform further prompt engineering model fine-tuning efforts, for example focusing on the inclusion of boundary cases or highlighting counterexamples for a particular class category.

Nevertheless, our AI-assisted live coding suffered from an inflated false positive rate, suggesting a respondent tendency toward acquiescence bias, where respondents confirmed incorrect labels more often than they should have, and a model tendency toward ‘over-classification’ where the model classified responses into categories even when no category clearly applied. This presents a design trade-off for researchers between reduced human coding effort and the risk of systematic over-classification. Researchers many consider implement more nuanced live-coding formats, for example forcing the model to re-code when confidence is low, or offering multiple candidate classifications and prompting the respondent to select all that apply. In settings where precision is critical, such as classification of low-incidence or sensitive behaviors, researchers may implement corrective strategies such as class-balanced fine-tuning or encoding prior expectations for class proportions directly within the zero-shot prompt.

For probing open-ended responses (RQ2), we first found that without any survey-specific training, the AI-enhanced chatbot was generally able to identify when elaboration was needed. We find support for our initial expectation that elaboration and relevance probes improve data quality in some ways (response specificity, explanatory details, and word length), but not others (response completeness, relevance, or linguistic variation). Moreover, effects on response quality were not uniform across different question types, suggesting that some survey tasks are better suited for open-ended elaboration (e.g., specifying a subcategory of occupation).

These findings carry practical implications for survey design. Before implementing AI-assisted probing at scale, researchers should identify which questions are at elevated risk of quality concerns such as low relevance, insufficient comprehensibility, or ambiguity from pretest or pilot data. Probing should be selectively applied where downstream analyses require precision or classification granularity and avoided for items where baseline performance is already strong.

On response experience (RQ3), among individuals who were recruited into our experiment, we find that the overall integration of AI into survey interviewing incurs a cost, albeit a small one. The very first probe in the elaboration/relevance condition led to slightly higher dropout than the equivalent point in the standardized condition, but those who remained reported only slightly less favorable experiences, even after receiving multiple probes by the end of the survey. Moreover, dropout rates were significantly lower than in previous evaluations of open-ended probing (Behr et al., 2012; Holland & Christian, 2009; Neuert & Lenzner, 2021). These results reinforce evidence that open-ended follow-ups, particularly when dynamically delivered, may not necessarily be disruptive, though they should still be used judiciously. Confirmation probing also had little impact on attrition or respondent experience and introduced only minimal shifts in response category distributions. It is perhaps unsurprising that confirmation probes have fewer negative impacts than other probes, as prior research indicates that close-ended (e.g., confirm vs. don't confirm) probes impose a lower cognitive burden than do open-ended probes (Neuert et al., 2023).

One practical recommendation that arises from our results is to administer open-ended probes sparingly, particularly for mobile users who are more sensitive to the respondent experience both in our study and previous research (De Bruijne & Wijnant, 2014). The effectiveness of probing did not appear to differ significantly between factual and opinion-based questions, though our evaluation is limited since only opinion-based questions included open-ended control responses for comparison.

## 6.2. Study Limitations

This study design involved several trade-offs. A key strength is its breadth: we evaluated AI-assisted conversational interviewing across multiple dimensions of data quality and respondent experience to create a more holistic view of the technology's implications. However, several limitations should be acknowledged.

Our implementation reflects performance using an "off-the-shelf" platform at a specific point in time. We view this as establishing a floor for current LLM capabilities, though future systems will likely perform better as the technology evolves. Due to constraints of our selected AI platform, elaboration and relevance probing were combined, preventing us from disentangling their separate effects. We also did not experimentally vary model prompts or parameter settings, which may influence data quality or respondent experience.

Respondents also knew they were interacting with an AI agent, which may not reflect future usage. It is also possible that responses do not represent human interactions with our interviewer if respondents themselves are engaging in 'AI-assisted responding' (Martherus et al., 2025; Westwood, 2025). Neither we, our platform vendor, nor our panel provider found any such indication of AI assistance. Lastly, our sample may not be representative of broader populations. While this study prioritizes internal validity over population-level inference, future researchers may wish to examine how these effects generalize through probability-based designs or careful sample re-weighting.

## 6.3. Future Directions

Despite promising results, researchers should apply LLMs in survey research with caution. As mentioned earlier, LLMs are fundamentally prone to hallucination in specific contexts even if 'factory evaluations' suggest such occurrences are minimal. While our conversational chatbots were not tasked with generating factual information, hallucinations in this context might involve ignoring or misapplying preprogrammed prompts used to guide probing behavior. As we have done in this study, researchers must extensively test any AI interviewing agents before taking them to the field, ensuring that they do not produce hallucinations that compromise data quality or respondent experience.

Additionally, just as complex or ambiguous survey questions can burden respondents, lengthy or vague answers may tax the chatbot, increasing response time and potentially reducing the quality of follow-up probes. Future studies should explore these limitations further by adjusting system parameters (e.g., temperature, tone), and testing different model configurations or types of models altogether.

LLMs are just one class of AI models capable of exhibiting human-like conversational behavior and alternatives may include small language models (SLMs) trained on more task-specific datasets or audio-based models designed for speech interaction. Trade-offs between speed, modality, task adaptability, and reasoning capacity will shape how well different AI agents perform in real-world data collection.

Improvements to the conversational interface itself might address problems of attrition, response bias, and poor respondent experience. Introducing ‘skip’ buttons into the chat interface could, for instance, convert attrition into non-response, providing a question-by-question lever for respondents to alleviate cognitive burden without exiting the survey. The acquiescence bias associated with ‘yes/no’ confirmation probes might be mitigated by displaying the chatbot’s coded category as the default or suggested choice in a categorical, rather than binary, confirmation probe. Researchers could also approach the process of live-coding using an entirely different mixed-format approach, where close-ended options are paired with an “other, specify” input, followed by confirmation probing only on “other specify” responses.

Finally, future studies might consider other AI-assisted interviewing techniques such as the dynamic generation of motivational statements or respondent feedback, both shown to increase respondent engagement in conventional probing (Dillman et al., 2008; Oudejans & Christian, 2010). Expanding beyond LLM-based ‘chatbots’ to audio-based conversational agents could be fruitful, given the differing information (Gavras et al., 2022) as well as rich context found in oral responses (Höhne et al., 2024) relative to written responses. Future work should carefully balance these benefits against potential costs in respondent experience. It is also important to recognize that these models are evolving rapidly. The benchmarks and evaluation metrics cited in this study, including human preference ratings, performance on truthfulness and toxicity benchmarks, and extensive pre-testing for our particular use case, justify our use of the SmartProbe model at the time of fielding (see Appendix A1). However, they may become outdated as new models with improved performance and safety characteristics become available. Researchers should therefore treat model selection as a moving target, continuously reassessing what constitutes “state-of-the-art” for a given task and context.

## Acknowledgments

This work was supported by a venture fund award from the NORC Business Ventures & Innovation unit. All research protocols in this study were approved by the institutional review board at NORC at the University of Chicago. Research subjects provided informed consent and confirmation of eligibility prior to participation. The authors would like to thank Skky Martin and Akari Oya for excellent research assistance. Additionally, we thank Ting Yan and Jeff Dominitz for feedback and advice.

# Appendix

## A1. Additional Experiment Details

### Figure A1. Probing Mechanism in Treatment Conditions

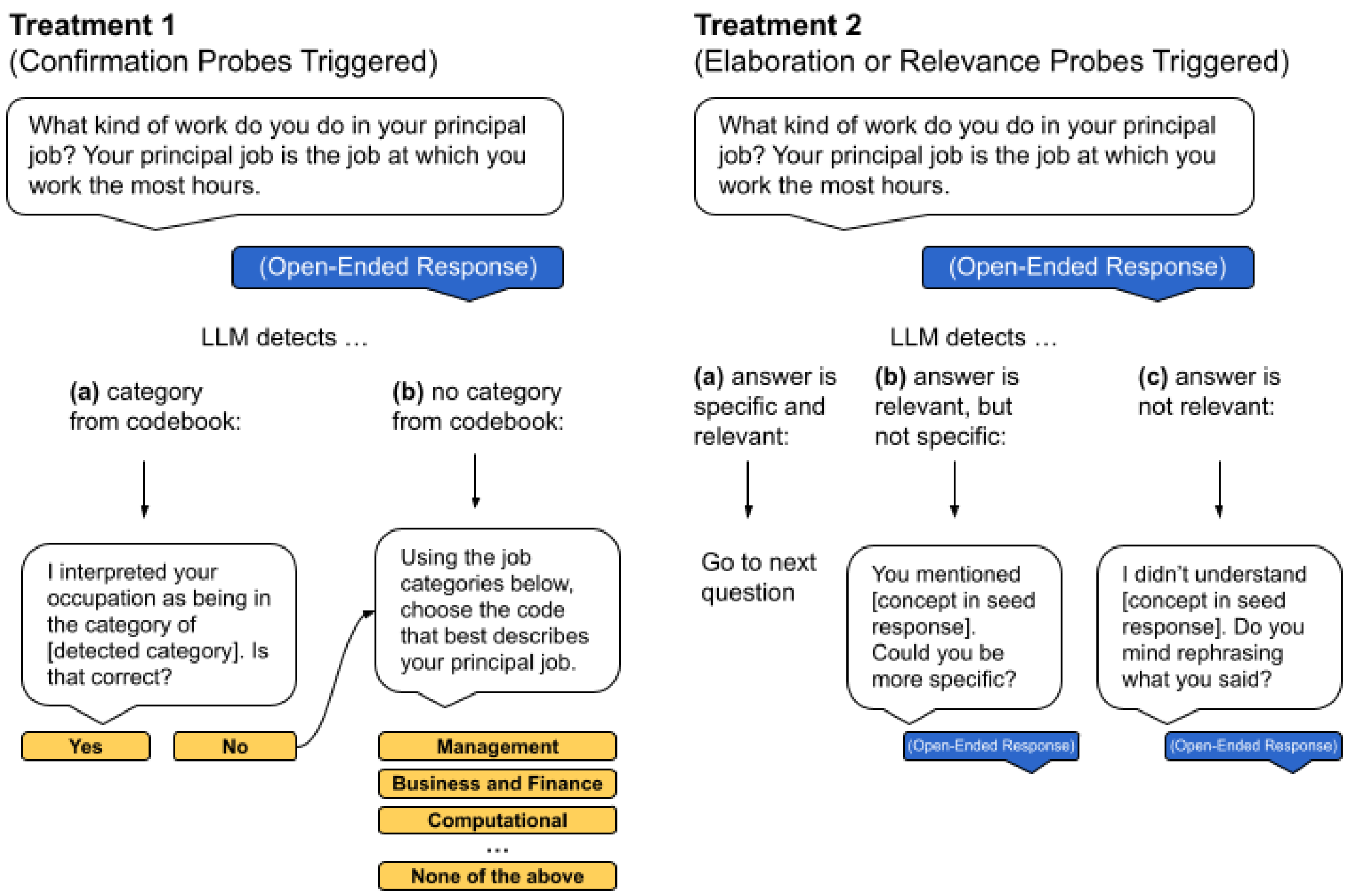

**Table A1. Full Questionnaire** (continued on following pages)

| **Q0.** Consent Screener | | |
|---|---|---|
| **Control**<br>(No Probes) | **Treatment 1**<br>(Confirmation Probes) | **Treatment 2**<br>(Elaboration or Relevance Probes) |
| In this study, researchers at a nonprofit research institution are interested in understanding the American public's opinions on issues of public importance. In addition to asking respondents about their attitudes on these issues, this survey will ask about demographic and employment characteristics in order to accurately create a representative profile of these attitudes.<br><br>Respondents may be randomly assigned to received follow-up questions generated by an AI chatbot. By selecting "I agree", you - as a participant in this study - agree to your responses being collected and analyzed as part of this study. You may refuse to participate by selecting "I do not agree."<br>1 I agree<br>2 I do not agree | | |

<table>
<tr><th colspan="3">Q1. Most Important Issue</th></tr>
<tr><th>Control<br>(No Probes)</th><th>Treatment 1<br>(Confirmation Probes)</th><th>Treatment 2<br>(Elaboration or Relevance Probes)</th></tr>
<tr><td colspan="3">[Seed Question] What do you think is the most important problem facing this country today? ________ (open-ended)</td></tr>
<tr><td>Go to next question</td><td>If LLM detects close-ended category in seed response:<br><br>➔ [Binary Confirmation Probe] I interpreted your answer as generally being about [sampled detected category]. Is that correct?<br>1 Yes<br>2 No<br><br>If response to binary confirmation probe is ‘No’ or LLM does not detect category in seed response:<br><br>➔ [Categorical Confirmation Probe] Which of the following topics did you mention in your answer to the previous question?<br>1 Economy<br>2 Cost of Living<br>3 Federal Budget<br>4 Jobs<br>5 Wages<br>6 Taxes<br>7 Economic Inequality<br>8 Corporate Power<br>9 International Trade<br>10 Immigration<br>11 Poverty<br>12 Elections and Democracy<br>13 Crime<br>14 Foreign Policy<br>15 Abortion<br>16 Race Relations and Racism<br>17 Climate Change<br>18 Education<br>19 Guns/Gun control<br>20 LGBTQ Issues<br>21 None of the above</td><td>If LLM detects that response is relevant, but not specific:<br><br>➔ [Elaboration Probe] (example) You mentioned [vague concept detected by LLM]. Could you tell me more?<br><br>Else if LLM detects that response is not relevant:<br><br>➔ [Relevance Probe] (example) I’m not sure what you mean by [irrelevant concept detected by LLM]. Could you rephrase what you said?<br><br>Else:<br><br>(Go to next question)</td></tr>
</table>

<table>
<tr><th colspan="3">Q2. Economic Conditions (Sentiment)</th></tr>
<tr><th>Control<br>(No Probes)</th><th>Treatment 1<br>(Confirmation Probes)</th><th>Treatment 2<br>(Elaboration or Relevance Probes)</th></tr>
<tr><td>[Seed Question] How would you rate economic conditions in this country today?<br>1 Very Good<br>2 Good<br>3 Neither Good Nor Bad<br>4 Bad<br>5 Very Bad</td><td colspan="2">[Seed Question] How would you rate the economic conditions in this country today?<br>______________ (open-ended)</td></tr>
<tr><td>Go to next question</td><td>If negative sentiment detected in seed response & positive sentiment not detected:<br><br>➔ [Binary Confirmation Probe] Just to confirm, are you saying that economic conditions are more negative than positive?<br>1 Yes<br>2 No<br><br>If positive sentiment detected in seed response & negative sentiment not detected:<br><br>➔ [Binary Confirmation Probe] Just to confirm, are you saying that economic conditions are more positive than negative?<br>1 Yes<br>2 No</td><td>If LLM detects that response is relevant, but not specific:<br><br>➔ [Elaboration Probe] (example) You mentioned [vague concept detected by LLM]. Could you tell me more?<br><br>Else if LLM detects that response is not relevant:<br><br>➔ [Relevance Probe] (example) I’m not sure what you mean by [irrelevant concept detected by LLM]. Could you rephrase what you said?<br><br>Else:<br><br>Go to next question</td></tr>
</table>

<table>
<tr><th colspan="3">Q3. Economic Conditions (Reason)</th></tr>
<tr><th>Control<br>(No Probes)</th><th>Treatment 1<br>(Confirmation Probes)</th><th>Treatment 2<br>(Elaboration or Relevance Probes)</th></tr>
<tr><td colspan="2">[Seed Question] What are the main reasons for your rating of economic conditions in this country?<br>_______________ (open-ended)</td><td rowspan="2">Triggered during Q2</td></tr>
<tr><td>Go to next question</td><td>If LLM detects close-ended category in seed response:<br><br>➔ [Binary Confirmation Probe] I interpreted your answer as generally being about [sampled detected category]. Is that correct?<br>1 Yes<br>2 No<br><br>If response to binary confirmation probe is ‘No’ or LLM does not detect category in seed response:<br><br>➔ [Categorical Confirmation Probe] Which of the following reasons matches your answer to the previous question?<br>1 Employment/Jobs<br>2 Layoffs<br>3 Inflation<br>4 Wages<br>(cont’d on next page)<br>5 Stock Market<br>6 Economic Growth<br>7 Government Spending<br>8 Gas Prices<br>9 Democratic / Biden Administration Policies<br>10 Interest Rates<br>11 Illegal immigration<br>12 Wealth inequality<br>13 Corporations or Corporate greed<br>14 Poverty or homelessness<br>15 Taxes<br>16 Politicians</td></tr>
</table>

| **Q4.** Preferred News Source | | |
|---|---|---|
| **Control**<br>(No Probes) | **Treatment 1**<br>(Confirmation Probes) | **Treatment 2**<br>(Elaboration or Relevance Probes) |
| **[Seed Question]** What is your main source of news?<br>1 Fox News or FoxNews.com<br>2 Local TV<br>3 CNN or CNN.com<br>4 Facebook<br>5 NBC<br>6 ABC or ABCNews.com<br>7 NPR<br>8 The New York Times<br>9 Local radio<br>10 CBS or CBSNews.com<br>11 MSNBC<br>12 The Washington Post<br>13 Another newspaper<br>14 Newsmax, OANN, Daily Wire or Daily Caller<br>15 Another TV network<br>16 Other | **[Seed Question]** What is your main source of news?<br>______ (open-ended) | |
| *Go to next question* | *If LLM detects close-ended category in seed response*:<br><br>➔ **[Binary Confirmation Probe]** I interpreted your answer as generally being about *[sampled detected category]*. Is that correct?<br>1 Yes<br>2 No<br><br>*If response to binary confirmation probe is 'No' or LLM does not detect category in seed response*:<br><br>➔ **[Categorical Confirmation Probe]** Do any of the following news sources match your previous answer?<br>1 Fox News or FoxNews.com<br>…<br>16 None of the above match my answer | *If LLM detects that response is relevant, but not specific*:<br><br>➔ **[Elaboration Probe]** (example) You mentioned *[vague concept detected by LLM]*. Could you tell me more?<br><br>*Else if LLM detects that response is not relevant*:<br><br>➔ **[Relevance Probe]** (example) I'm not sure what you mean by *[irrelevant concept detected by LLM]*. Could |
| | | you rephrase what you said?<br><br>*Else:*<br><br>*Go to next question* |

| Demographics Module | | | |
|---|---|---|---|
| **Q#** | **Control**<br>(No Probes) | **Treatment 1**<br>(Confirmation Probes) | **Treatment 2**<br>(Elaboration or Relevance Probes) |
| **Q5** | **[Age]** Thank you for answering those questions. Now I have a few questions about you…<br><br>What is your age?<br>___ (numeric input) | | |
| **Q6** | **[Gender]** How would you describe your gender?<br>1 Male<br>2 Female<br>3 Other | | |
| **Q7** | **[Educ]** What is the highest level of education you have completed?<br>1 Less than high school<br>2 High school graduate or equivalent<br>3 Some college/associate's degree<br>4 Bachelor's degree<br>5 Postgraduate study/professional degree | | |
| **Q8** | **[Employment]** Are you currently working?<br>1 Yes<br>2 No | | |

<table>
<tr><th colspan="3">If response to Q8 is Yes:<br>Q9. Main Occupation</th></tr>
<tr><th>Control<br>(No Probes)</th><th>Treatment 1<br>(Confirmation Probes)</th><th>Treatment 2<br>(Elaboration or Relevance Probes)</th></tr>
<tr><td>[Seed Question] What kind of work do you do in your principal job? Your principal job is the job at which you work the most hours. Select from the following list of job codes from the Bureau of Labor Statistics that best describes your principal job.<br>1 Management Occupations<br>2 Business and Financial Operations Occupations<br>3 Computer and Mathematical Occupations<br>4 Architecture and Engineering Occupations<br>5 Life, Physical, and Social Science Occupations<br>6 Community and Social Service Occupations<br>7 Legal Occupations<br>8 Educational Instruction and Library Occupations<br>9 Arts, Design, Entertainment, Sports, and Media Occupations<br>10 Healthcare Practitioners and Technical Occupations<br>11 Healthcare Support Occupations<br>12 Protective Service Occupations<br>13 Food Preparation and Serving Related Occupations<br>14 Building and Grounds Cleaning and Maintenance Occupations<br>15 Personal Care and Service Occupations<br>16 Sales and Related Occupations<br>17 Office and Administrative Support Occupations<br>18 Farming, Fishing, and Forestry Occupations<br>19 Construction and Extraction Occupations<br>20 Installation, Maintenance, and Repair Occupations<br>21 Production Occupations<br>22 Transportation and Material Moving Occupations<br>23 Military Specific Occupations</td><td colspan="2">[Seed Question] What kind of work do you do in your principal job? Your principal job is the job at which you work the most hours.<br>_____ (open-ended)</td></tr>
<tr><td>Go to next question</td><td colspan="2">If LLM detects close-ended category in seed response:<br><br>➔ [Binary Confirmation Probe] I interpreted your answer as generally being about [sampled detected category]. Is that correct?<br>1 Yes<br>2 No<br><br>If response to binary confirmation probe is ‘No’ or LLM does not detect category in seed response:</td></tr>
</table>

| | |
|---|---|
| | ➔ **Categorical Confirmation Probe:** Which of the following reasons matches your answer to the previous question?<br>1 Management Occupations<br>…<br>24 None of the above |

| Respondent Experience Module | | | |
|---|---|---|---|
| Q# | Control (No Probes) | Treatment 1 (Confirmation Probes) | Treatment 2 (Elaboration or Relevance Probes) |
| Q10 | Finally, please answer the following questions about your survey experience.<br><br>**[Tone]** Overall, how formal was the tone of questions in this survey?<br>1 Very formal<br>2 Somewhat formal<br>3 Not very formal<br>4 Not formal at all | | |
| Q11 | **[Quality]** Overall, how would you rate the quality of your responses to questions in this survey?<br>1 Very high quality<br>2 Somewhat high quality<br>3 Somewhat low quality<br>4 Very low quality | | |
| Q12 | **[Ease]** How easy was it to complete this survey?<br>1 Very easy<br>2 Somewhat easy<br>3 Neither easy nor difficult<br>4 Somewhat difficult<br>5 Very difficult | | |
| Q13 | **[Frustration]** How frustrating was this survey experience?<br>1 Very frustrating<br>2 Somewhat frustrating<br>3 Not very frustrating<br>4 Not frustrating at all | | |

Our survey experiment was fielded on the conversational AI platform Inca, with an interface structurally identical to that depicted in Figure 1 in the manuscript. The large language model (LLM) used to administer probes and conduct classification in our experiment is SmartProbe (Seltzer et al., 2023), a model fine-tuned from the InstructGPT family (Ouyang et al., 2022) which are a collection of models themselves fine-tuned from the GPT-3 foundational model (Brown et al., 2020).

**Structure.** To generate probes in a real-time survey, the model is given an instructional prompt consisting of (1) the seed question, (2) the respondent's response, along with (3) structural information about the dialogue context—such as whether the prior response was vague, off-topic, or already specific—provided by a lightweight semantic classifier trained to detect relevance and elaboration needs.[7] This classifier helps the system determine whether a follow-up probe should seek clarification or elaboration, enabling the prompt to convey both the content and conversational intent of the prior exchange. Multiple candidate probes are generated, screened for clarity, safety, and contextual relevance, and then ranked. If none pass a quality threshold, a rule-based backup probe was used.

While the exact prompt format used by the proprietary SmartProbe system is not publicly available, the following simplified structure reflects the key components used to generate probing questions:

```
Seed Question: "What is your current occupation?"
Respondent Answer: "I work in a lab."

[Dialogue Context]:
- Answer Type: Short
- Relevance: Relevant
- Specificity: Low
- Requires Elaboration: Yes

[Optional Metadata]:
- Target Codebook Category: Medical or Scientific Occupations

Instruction: Generate a follow-up question that asks the respondent to
elaborate on their job role using clear and concise language.
```

**Fine Tuning and Evaluation.** InstructGPT models, upon which our LLM was based, were trained using a multi-step fine-tuning process designed to align language model behavior with human preferences (Ouyang et al., 2022). This process began by collecting human-written examples of desired outputs in response to user prompts. Human annotators were then asked to compare different model outputs and indicate which they prefer. These preferences are used to train the model to better follow instructions and produce high-quality responses. The resulting models are evaluated both by human raters and on standardized benchmarks. For instance, on the `TruthfulQA` benchmark – a set of questions designed to test whether models give factually accurate answers – InstructGPT produced truthful responses about twice as often as GPT-3. On the `RealToxicityPrompts` dataset – a collection of prompts used to assess how often models generate harmful or offensive content – InstructGPT reduced toxic completions by approximately 25% when prompted to be respectful. Notably, human raters preferred outputs from the smaller InstructGPT model (1.3 billion parameters) over the much larger GPT-3 model (175 billion parameters) in 70–85% of comparisons, demonstrating the effectiveness of fine-tuning.

[7] As SmartProbe is a proprietary system developed by the vendor, the exact instructional prompts used to generate probes are not publicly available.

SmartProbe built on this foundation through additional fine-tuning on domain-specific examples from the vendor's proprietary Market Research Knowledge Base. This corpus includes question–answer pairs written by professional researchers across topics such as advertising testing, brand perception, and customer experience. In prior vendor evaluations, 69% of SmartProbe-generated probe questions – elicited via the augmented seed question/response/dialogic context prompts described above – were rated 4 or 5 (on a 5-point scale) by expert annotators, with fewer than 1% rated as poor quality. In a parallel field test with over 900 North American respondents, 76% of responses elicited by SmartProbe were rated high quality, compared to only 25% for responses to standard, generic probes.

**Deployment.** In our study, SmartProbe was deployed with a temperature setting of 0 to ensure reproducibility and a relatively high presence penalty to discourage verbatim copying of training examples.

## Table A2. Completions by Treatment Condition

| Condition | Completes | Dropouts |
|---|---|---|
| **Control** | 601 | 43 |
| **Treatment 1 (Conf. Probing)** | 601 | 64 |
| **Treatment 2 (Elab/Rel. Probing)** | 601 | 88 |

## Table A3. Summary Statistics of Interview Duration by Treatment Condition

| Condition | Duration of Interview (Minutes) | | | | | |
|---|---|---|---|---|---|---|
| | **1%** | **25%** | **Median** | **Mean** | **75%** | **99%** |
| **Control** | 0m | 1.9m | 2.4m | 3.1m | 3.2m | 14.2m |
| **Treatment 1 (Conf. Probing)** | 0m | 2.5m | 3.3m | 4.1m | 4.7m | 15.8m |
| **Treatment 2 (Elab/Rel. Probing)** | 0m | 3.1m | 4.5m | 5.9m | 6.8m | 29.3m |

## Table A4. Summary Statistics of Completion Time by Treatment Condition

| Condition | Duration of Interview Among Non-Dropouts (Minutes) | | | | | |
|---|---|---|---|---|---|---|
| | **1%** | **25%** | **Median** | **Mean** | **75%** | **99%** |
| **Control** | 1.3m | 1.9m | 2.5m | 3.3m | 3.3m | 16.1m |
| **Treatment 1 (Conf. Probing)** | 1.8m | 2.7m | 3.5m | 4.4m | 4.8m | 16.2m |
| **Treatment 2 (Elab/Rel. Probing)** | 1.9m | 3.4m | 4.9m | 6.4m | 7.1m | 29.3m |

## Figure A1. Sample Composition

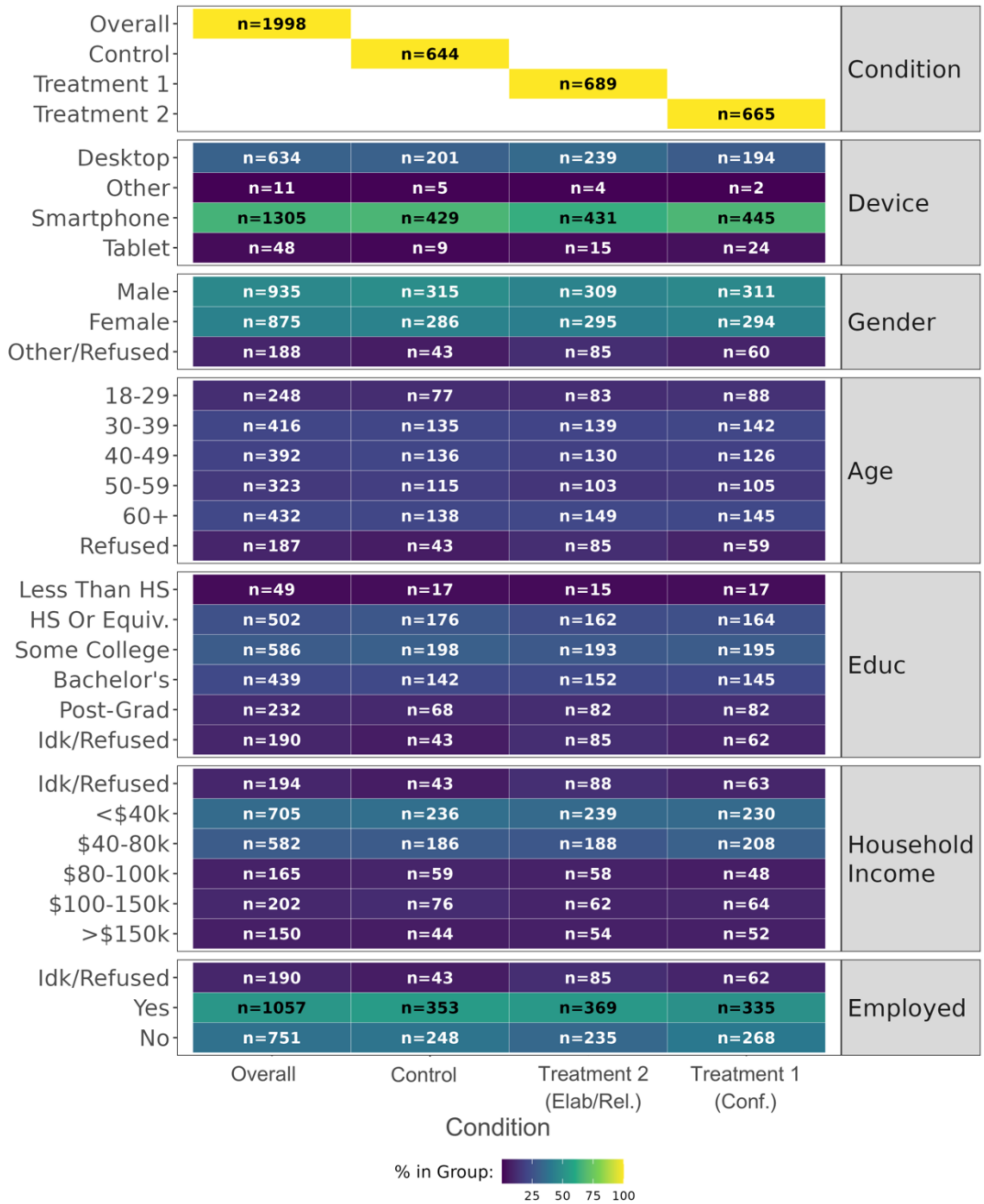

**Figure A2. Summary of Human-Coded Quality Criteria**

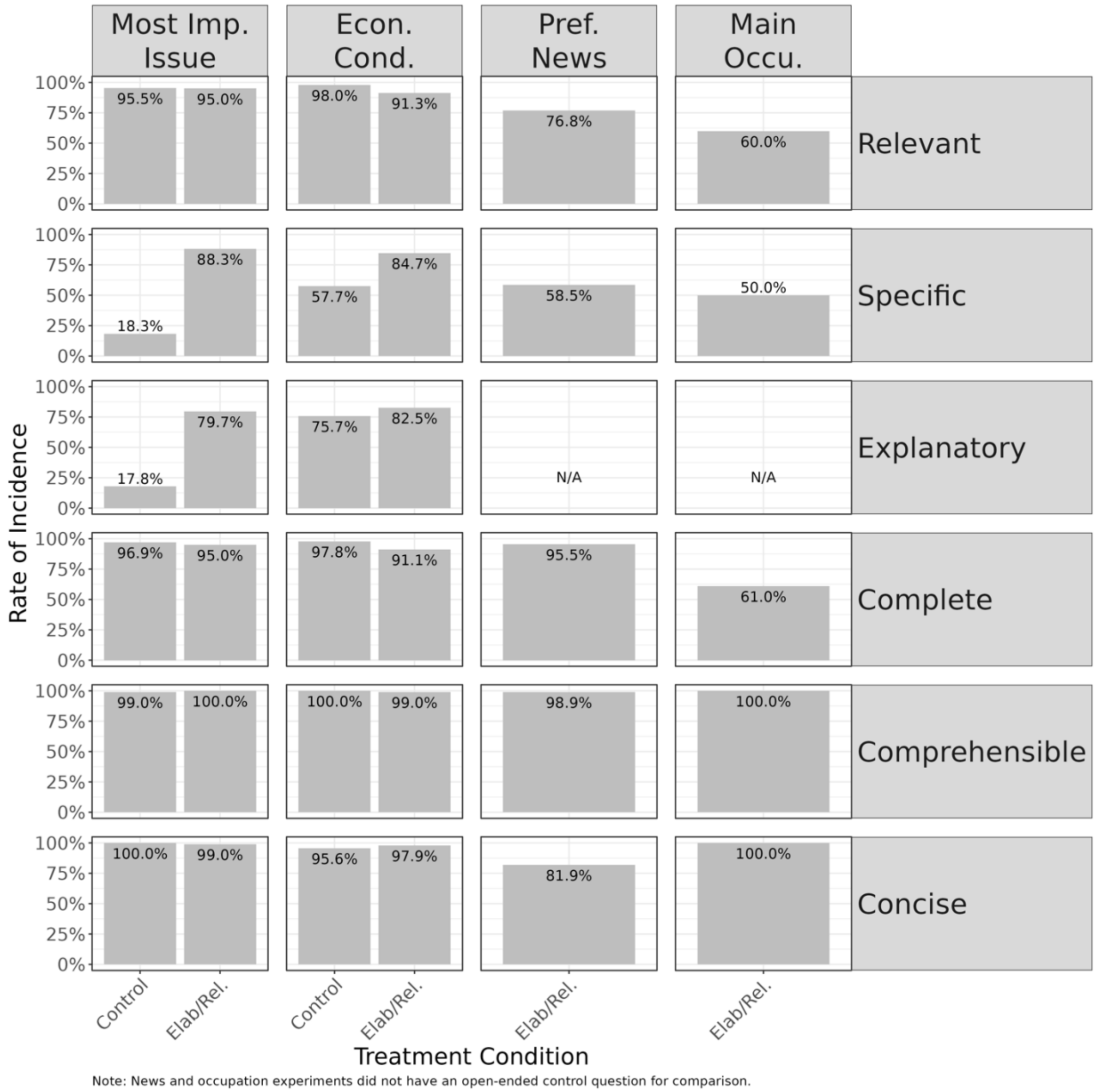

**Figure A3. Summary of NLP-Based Informational Measures**

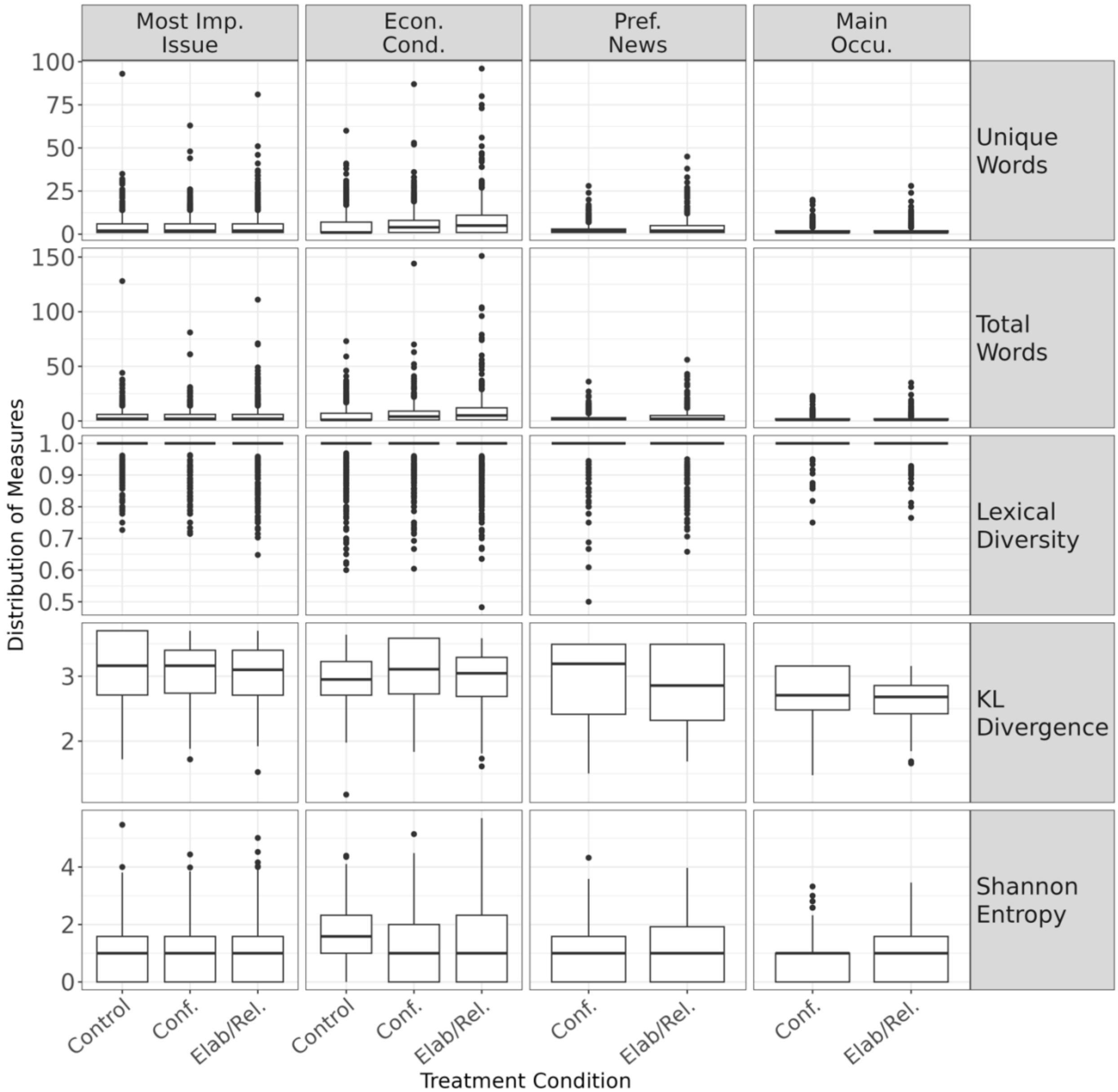

Note: Measures are shown for the post-probing response text in the elab/rel. treatment condition

**Figure A4. Summary of Respondent Attrition**

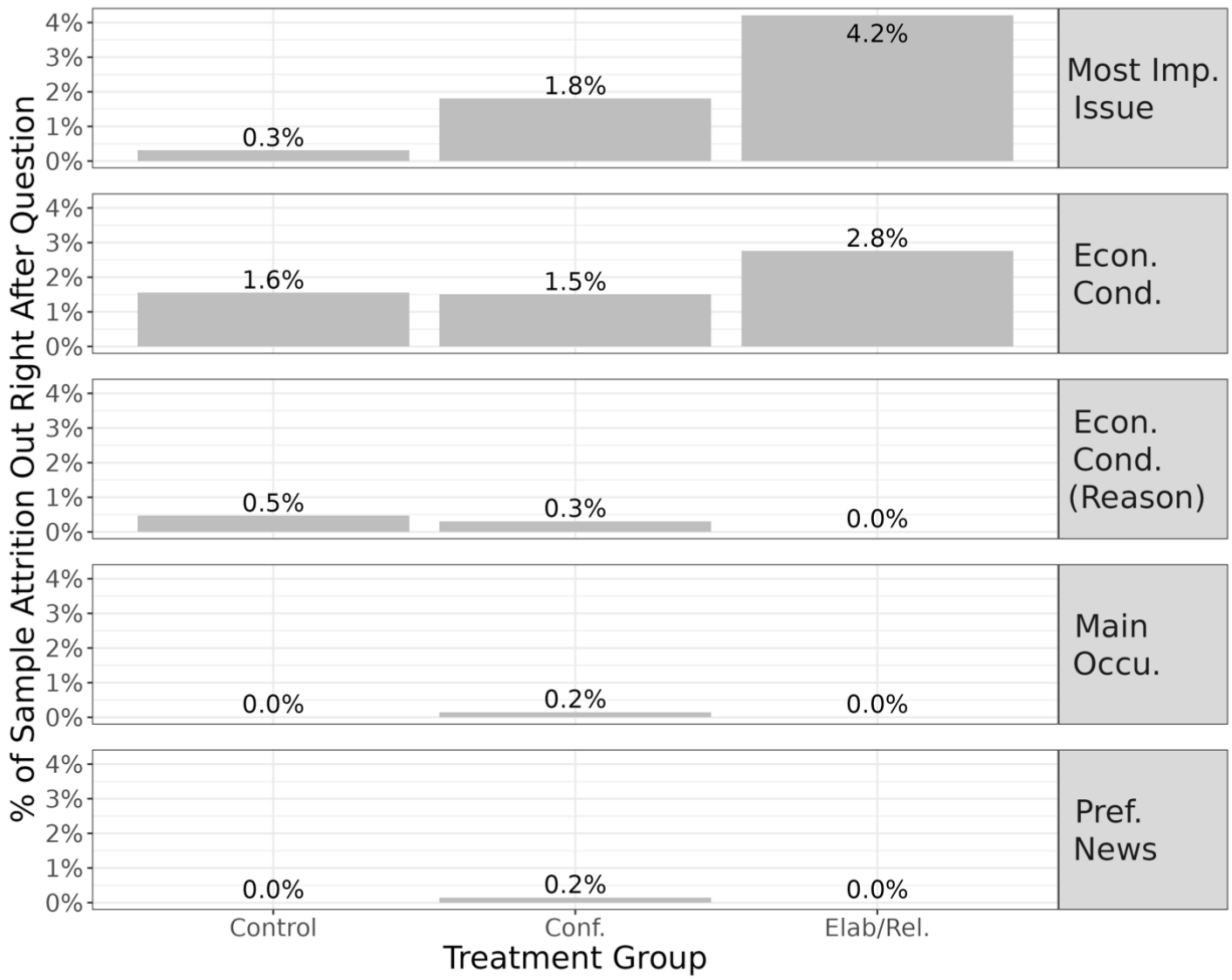

**Figure A5. Summary of Self-Reported Respondent Experience**

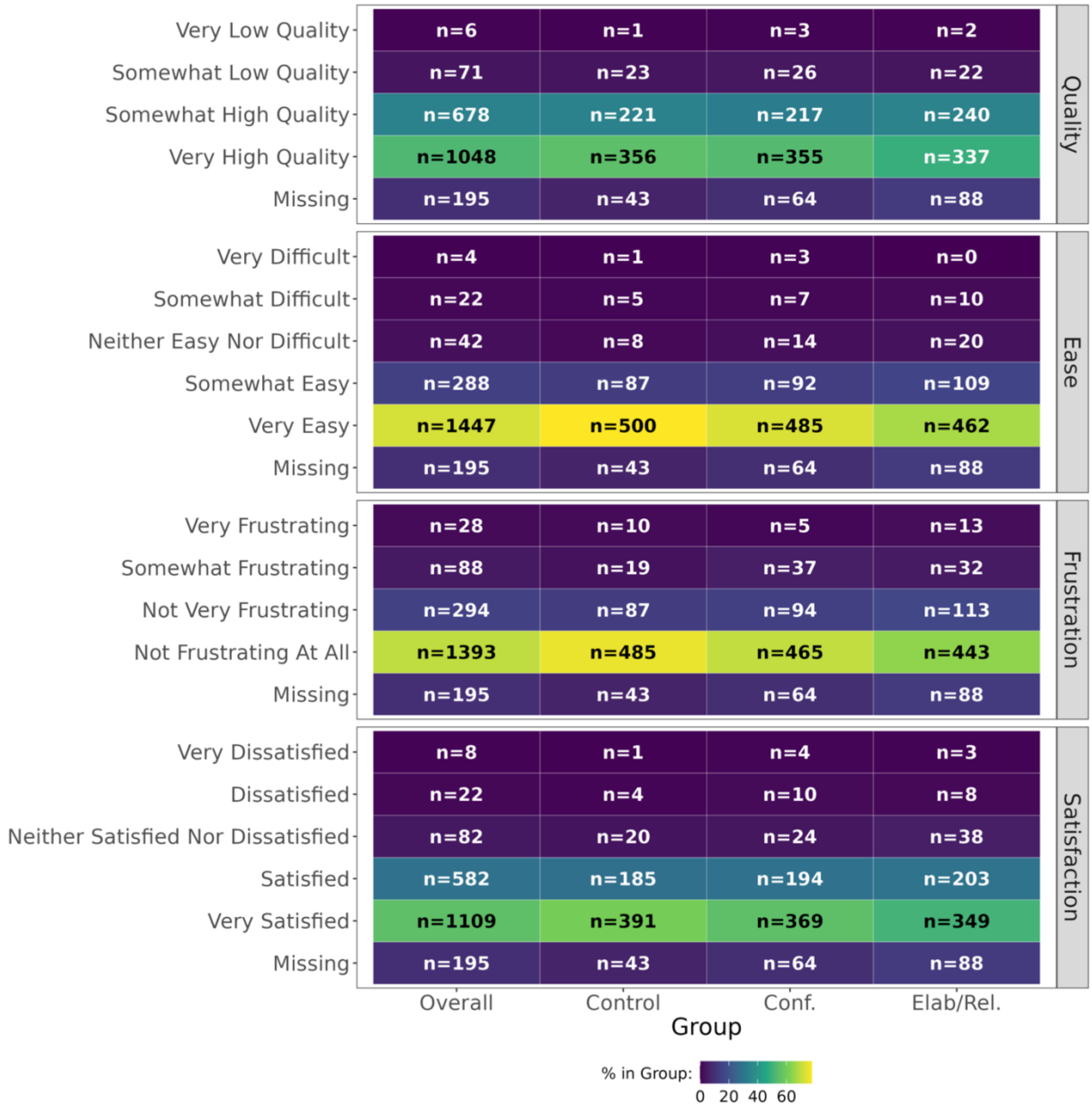

# A2. Additional Live Coding Results

### Figure A6. Evidence of Acquiescence Bias in Confirmation Probing: Comparison of Response Categories (Coded by Coders in Treatment 1 vs. Confirmed by Respondent in Treatment 1)

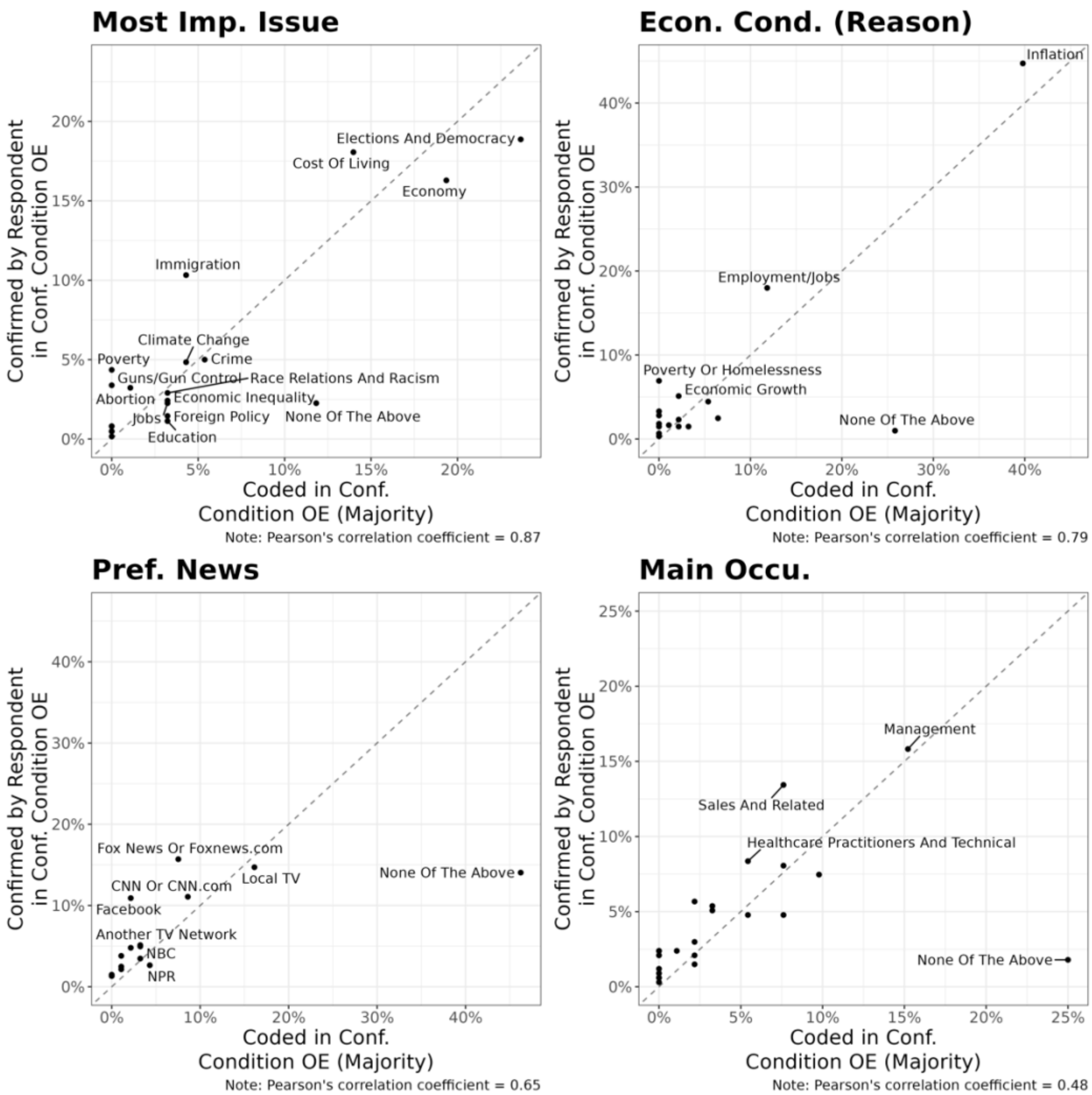

**Figure A7. Comparison of Response Categories (Confirmed in Treatment 1 vs. Coded By Coders in Other Conditions)[8]**

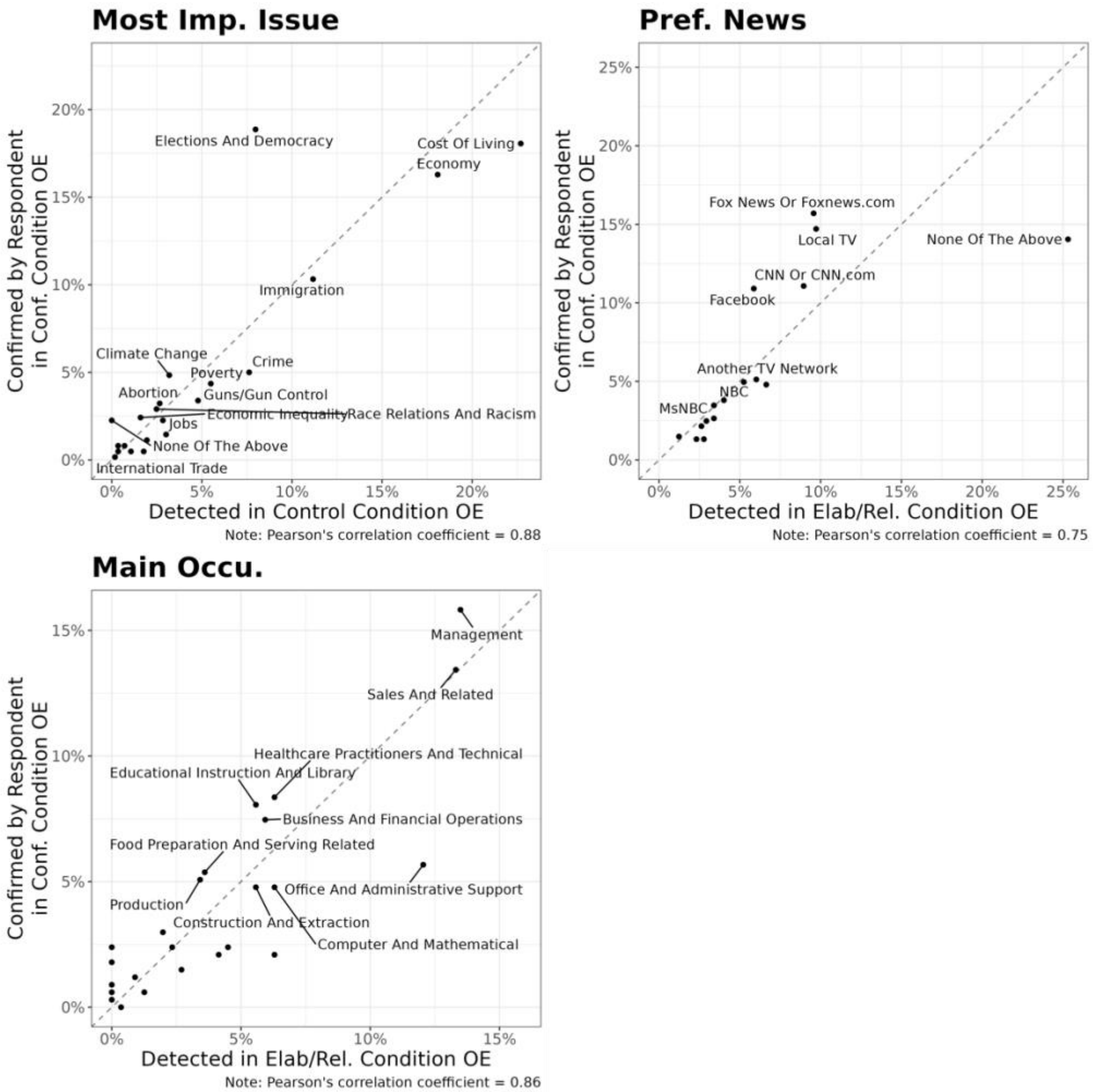

[8] Due to the way the economic conditions (reason) question was asked in the control condition (close-ended) versus the elaboration/relevance conditions (as a probe, rather than a standalone static question), coding of categories was not possible and is therefore omitted from Figure 4.

**Figure A8. Comparison of Response Categories (Live Coded by Chatbot in Treatment 1 vs. Answered in Control Close-End)**[9]

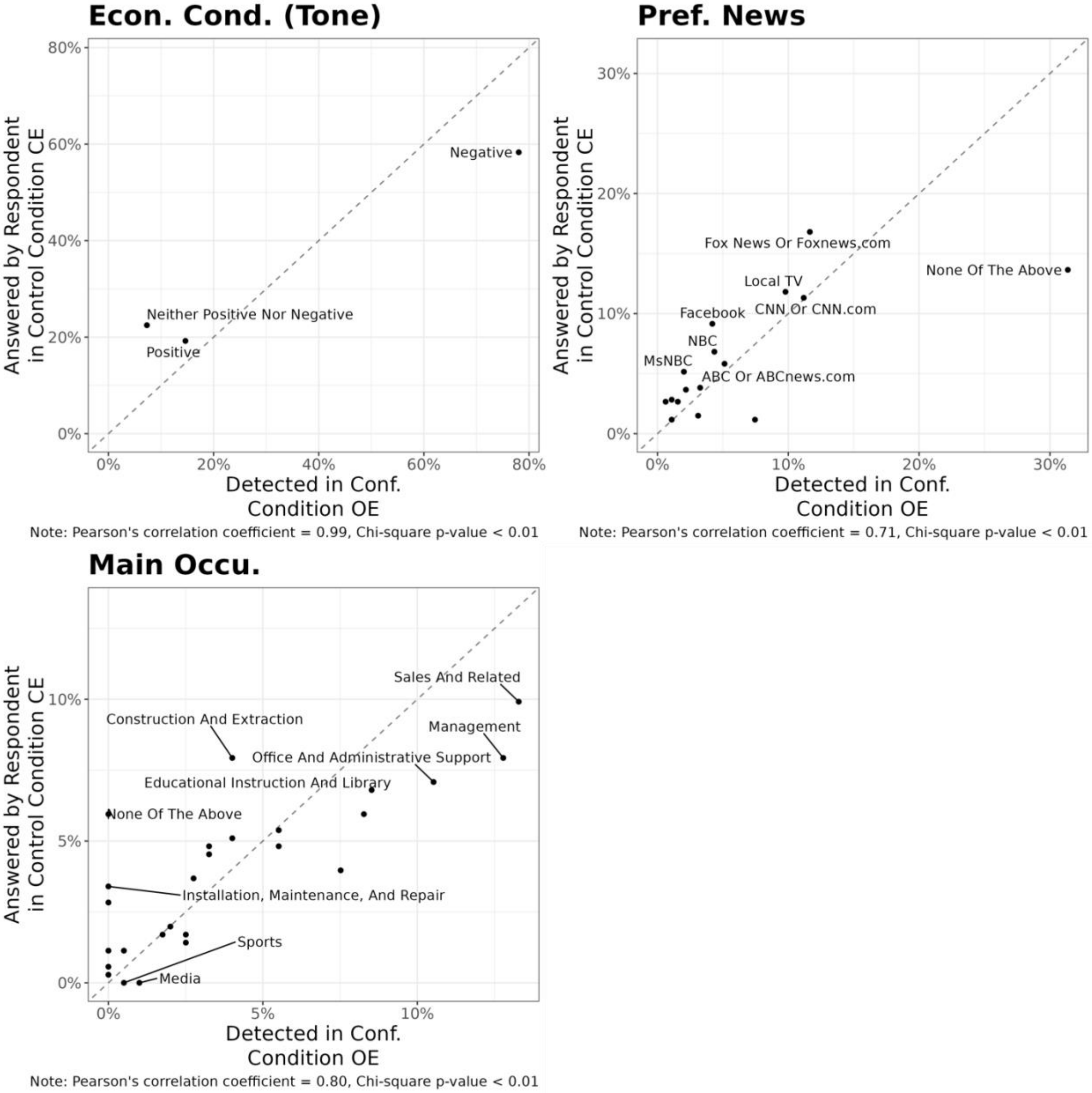

[9] "Tone" here is used interchangeably with "sentiment".

**Figure A9. Comparison of Response Categories (Live Coded by Chatbot in Treatment 1 vs. Coded in Other Open-Ended Conditions)**

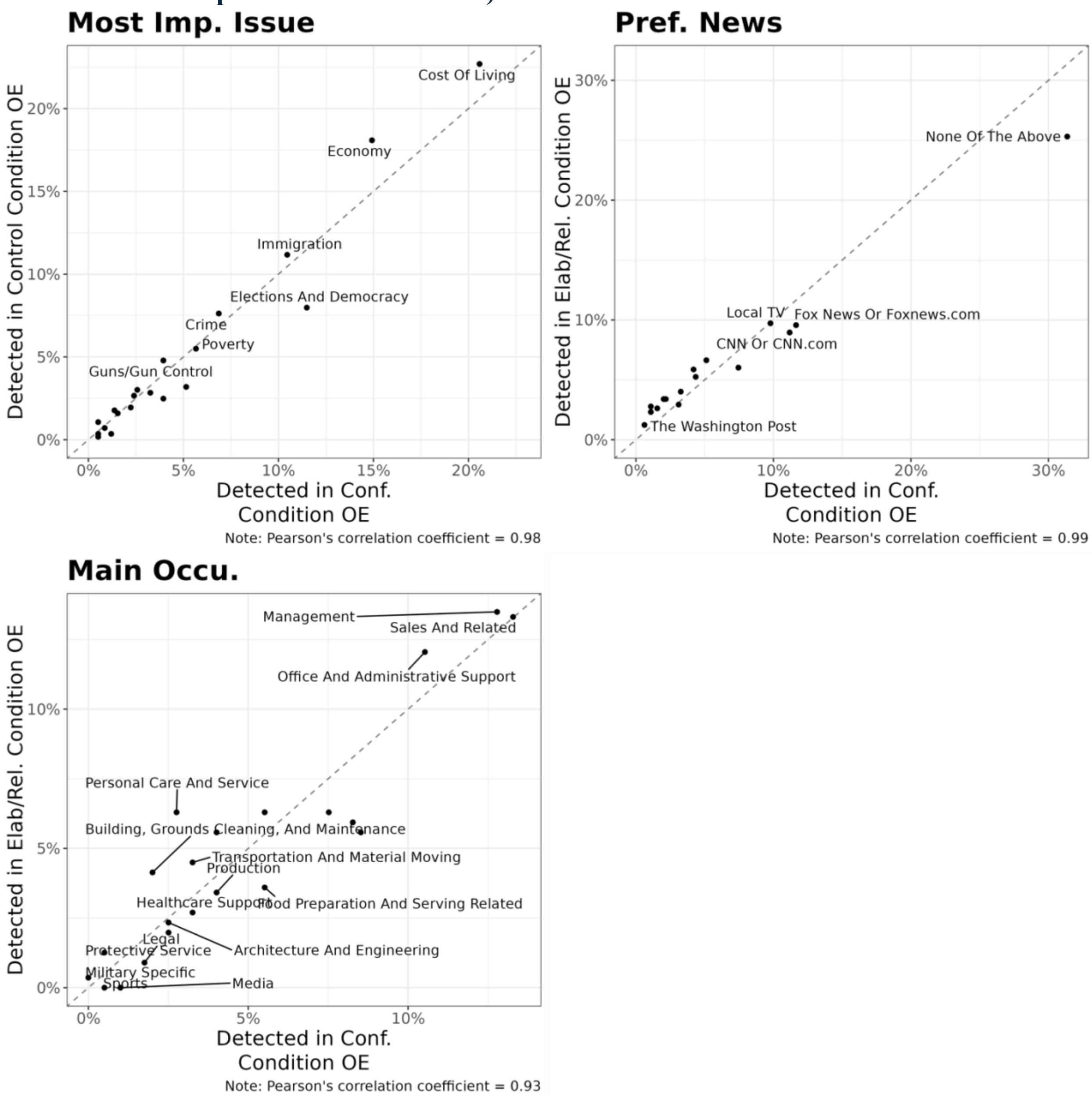

**Figure A10. Comparison of Response Categories (Confirmed in Treatment 1 vs. Answered in Control Close-End)**

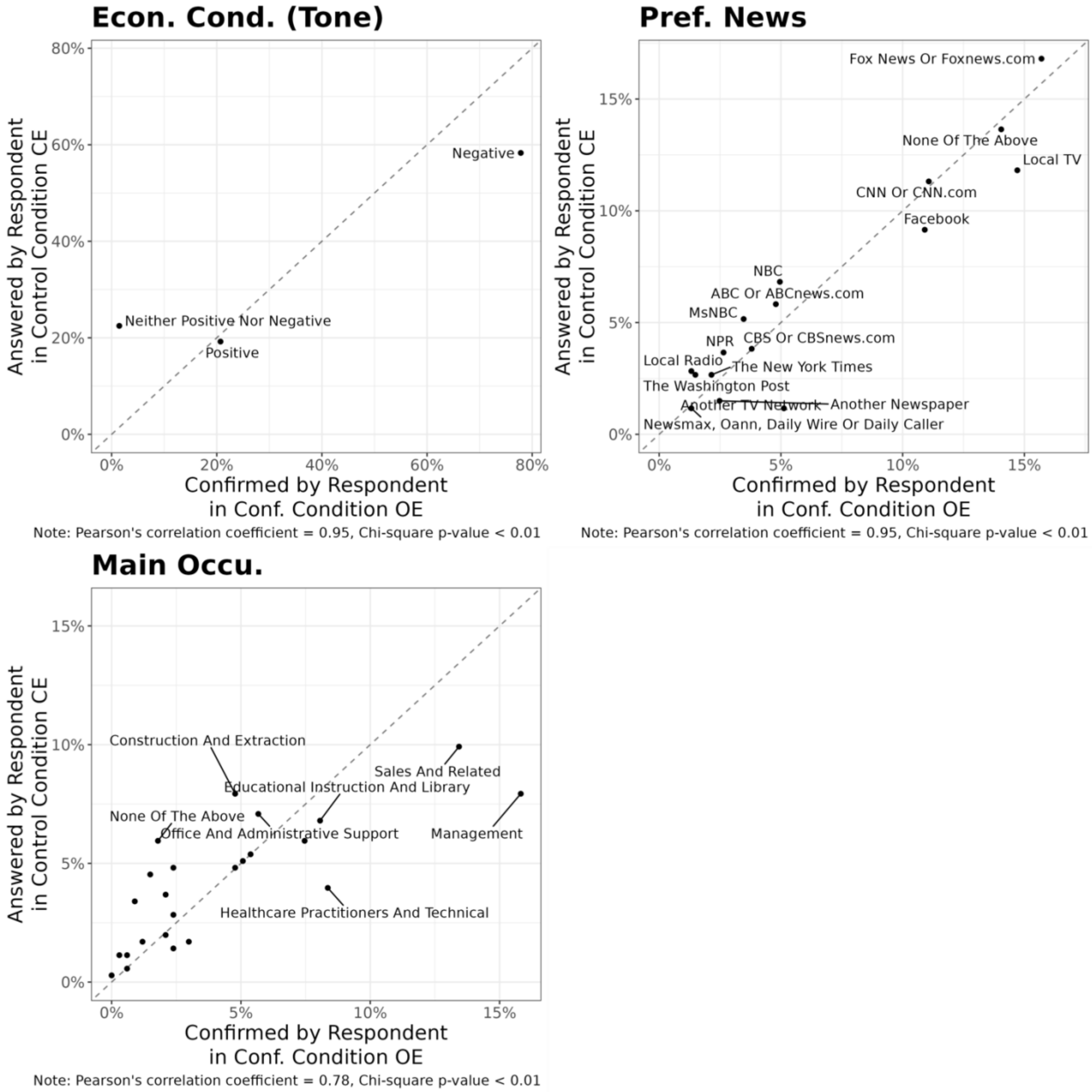

## Figure A11. Comparison of Response Categories (Confirmed by Respondent in Treatment 1 vs. Live Coded by Chatbot)

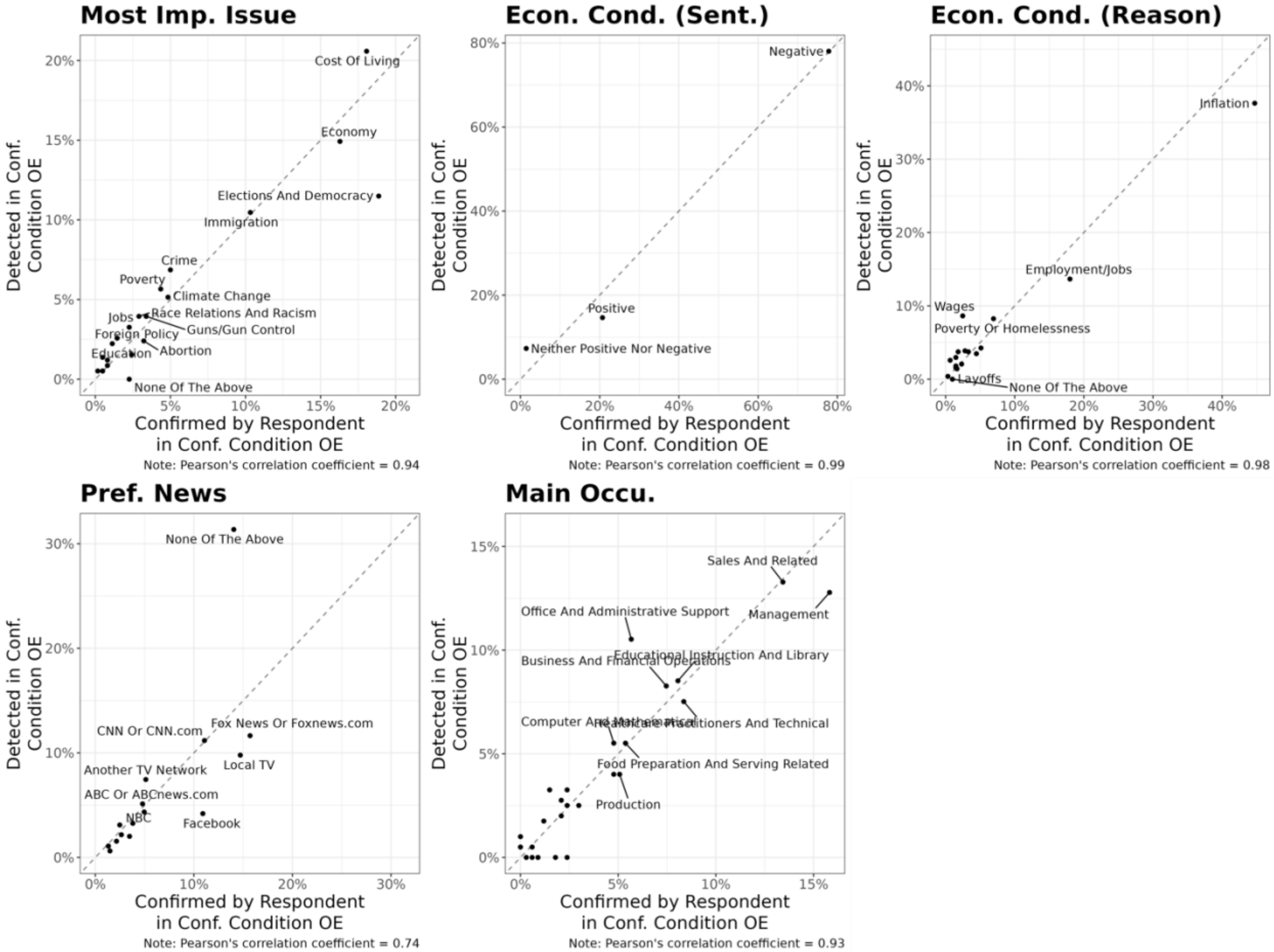

## Table A5. Respondent Confirmation (Treatment 1) by Device Type

| Device | Most Imp. Issue | Econ. Cond. (Sentiment) | Econ. Cond. (Reason) | Pref. News | Main Occu. |
|---|---|---|---|---|---|
| **Smartphone** (n=445) | 74.0% | 95.9% | 77.1% | 62.8% | 87.7% |
| **Desktop** (n=194) | 73.3% | 97.2% | 88.8% | 74.0% | 76.8% |
| **Tablet** (n=24) | 69.6% | 95.0% | 91.3% | 65.2% | 90.9% |

**Figure A12. Live Coding (Treatment 1) Confusion Matrices**

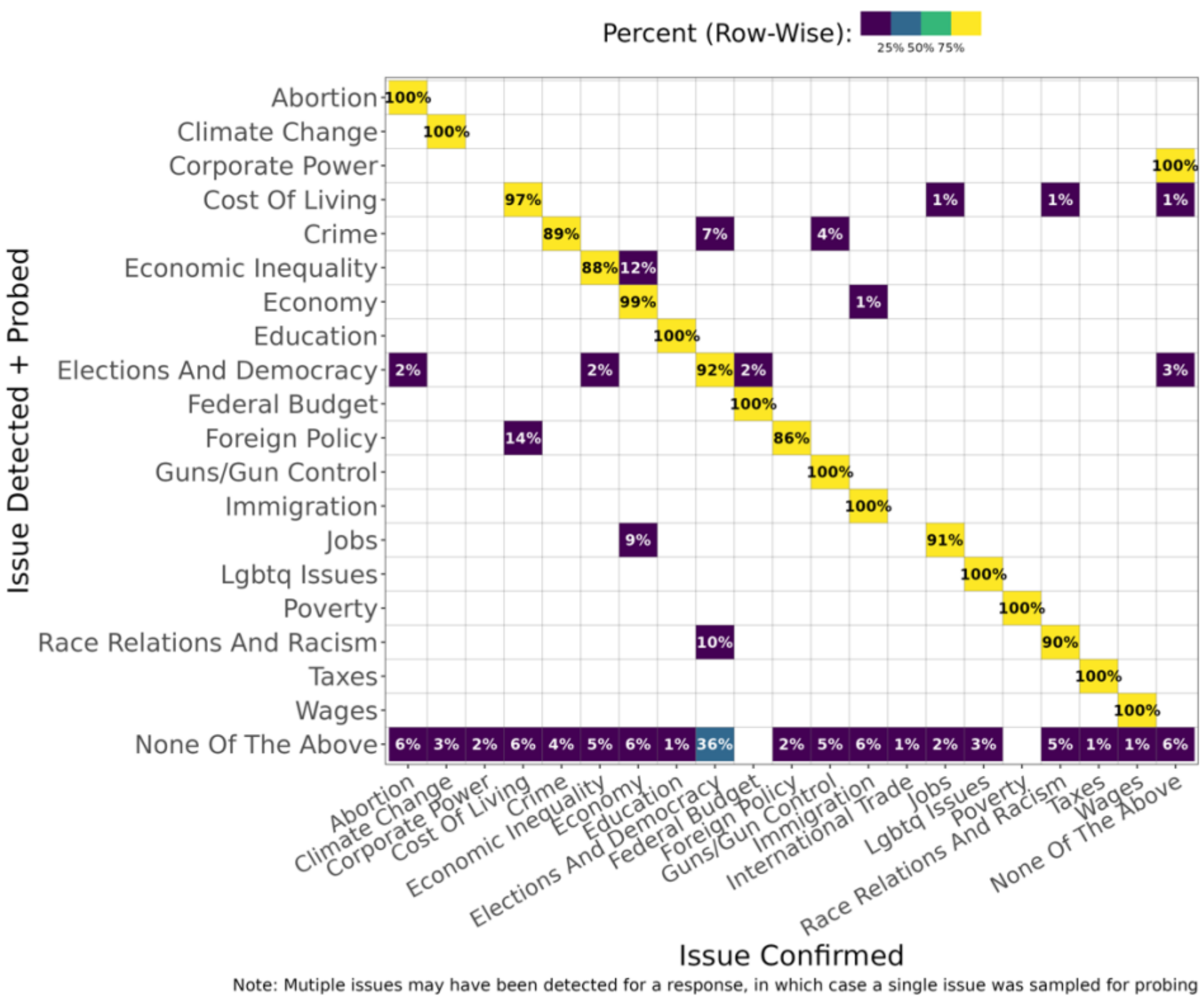

**(a) Most Important Issue**

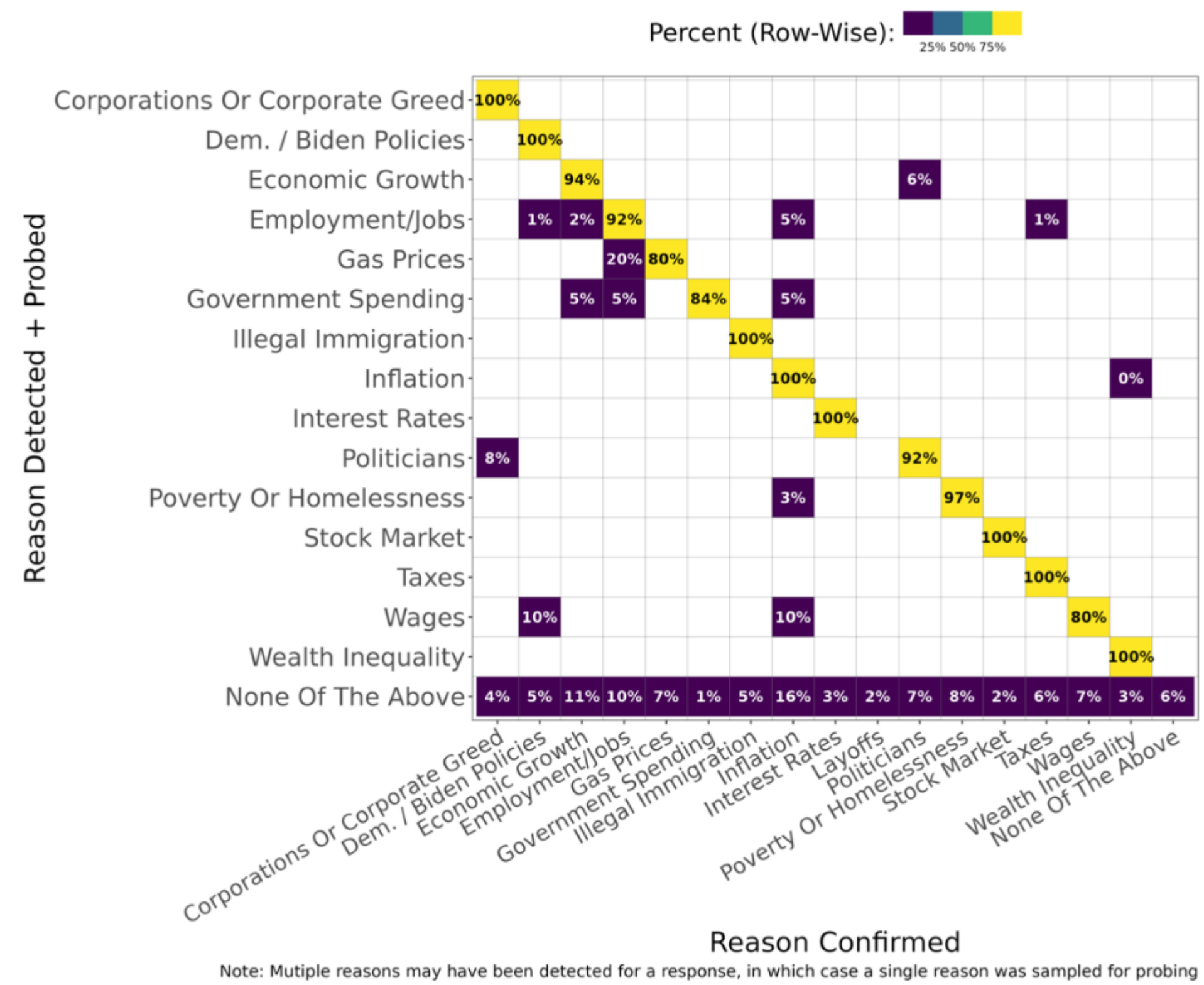

**(b) Economic Conditions (Reason)**

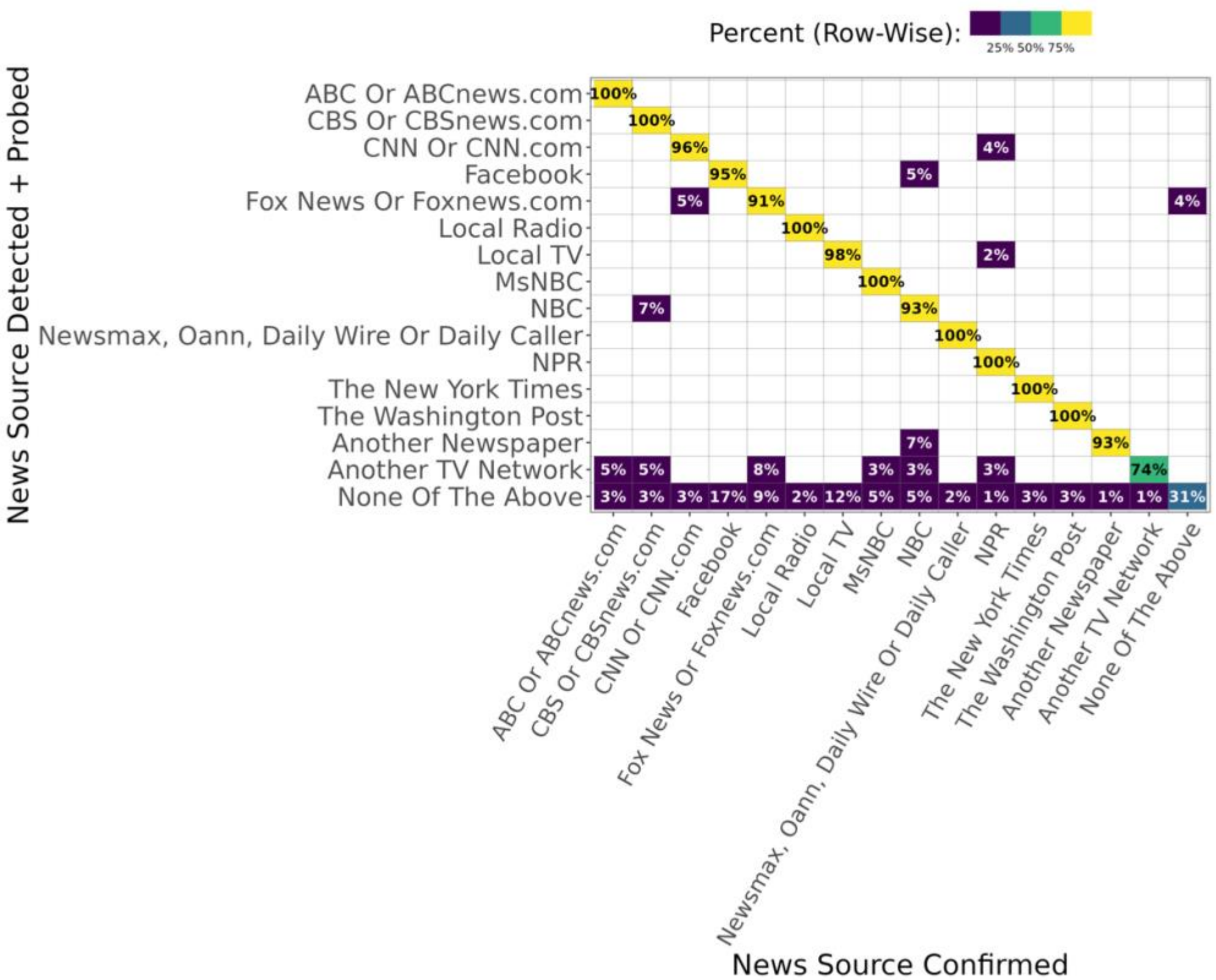

(c) Preferred News Source

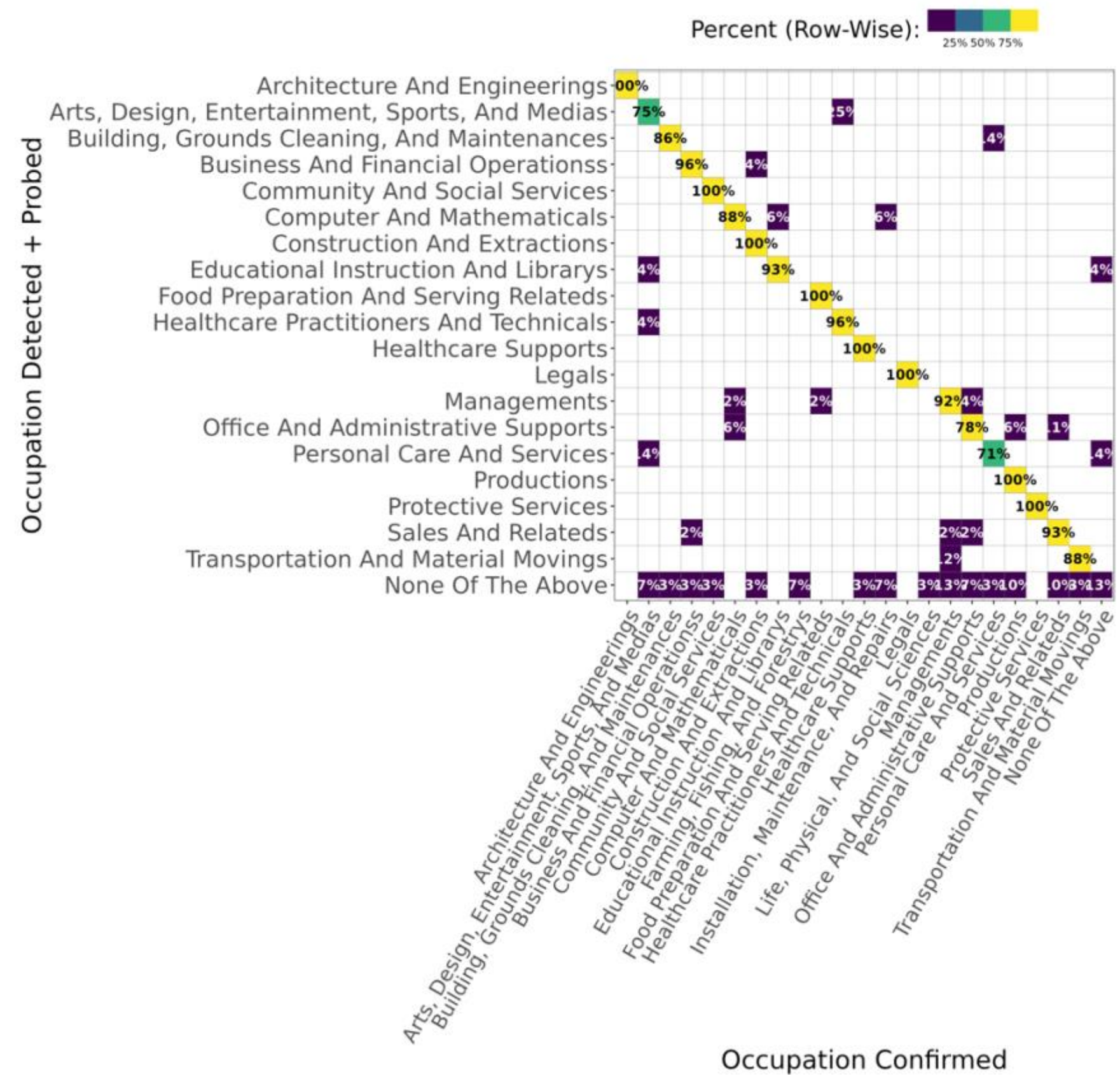

(d) Occupation

# A3. Additional Probing Results

Regression results in this section follow the same corrections and specifications as in the main text.

## Table A6. Examples of Elaboration/Relevance Probes

| Probe Triggered | Example Question Sequence | | | |
|---|---|---|---|---|
| | **Most Imp. Issue** | **Econ. Cond.** | **Pref. News** | **Main Occu.** |
| **Elaboration Probe** | **[Seed Response]** Economy and inflation<br><br>**[Chatbot Probe]** Thank you for sharing your thoughts. Could you please elaborate on why you think the economy and inflation are the most important problems facing the country today? | **[Seed Response]** Our food bank needs more donations since more people are in need<br><br>**[Chatbot Probe]** It sounds like you're saying that there are more people in need of assistance. Can you tell me more about how the economy is affecting people's ability to provide for themselves? | **[Seed Response]** New York Times and the Washington Post.<br><br>**[Chatbot Probe]** That's interesting. Can you tell me more about why you prefer these sources over others? | **[Seed Response]** Contractor<br><br>**[Chatbot Probe]** That's interesting. Could you tell me more about the type of contracting work you do? |
| **Relevance Probe** | **[Seed Response]** Democracy not what religion<br><br>**[Chatbot Probe]** I'm sorry, I'm not sure I understand your response. Could you please elaborate on what you mean by "democracy not what religion"? | **[Seed Response]** good<br><br>**[Chatbot Probe]** Hmm I think I know what you mean, but could you explain a bit more? | **[Seed Response]** Internet and my husband<br><br>**[Chatbot Probe]** Thank you for sharing. Could you please clarify which specific websites or online platforms you use to consume news? And when you say your husband, do you mean he shares news articles with you or do you discuss news topics together? | **[Seed Response]** per week 35 hours<br><br>**[Chatbot Probe]** I'm sorry, but I think there might have been a misunderstanding. I was asking about the type of work you do in your principal job. Could you please provide more information about the kind of work you do? |

**Figure A13. Comparison of Seed Response Quality (Treatment 2 Probe Triggered vs. No Probe Triggered)**

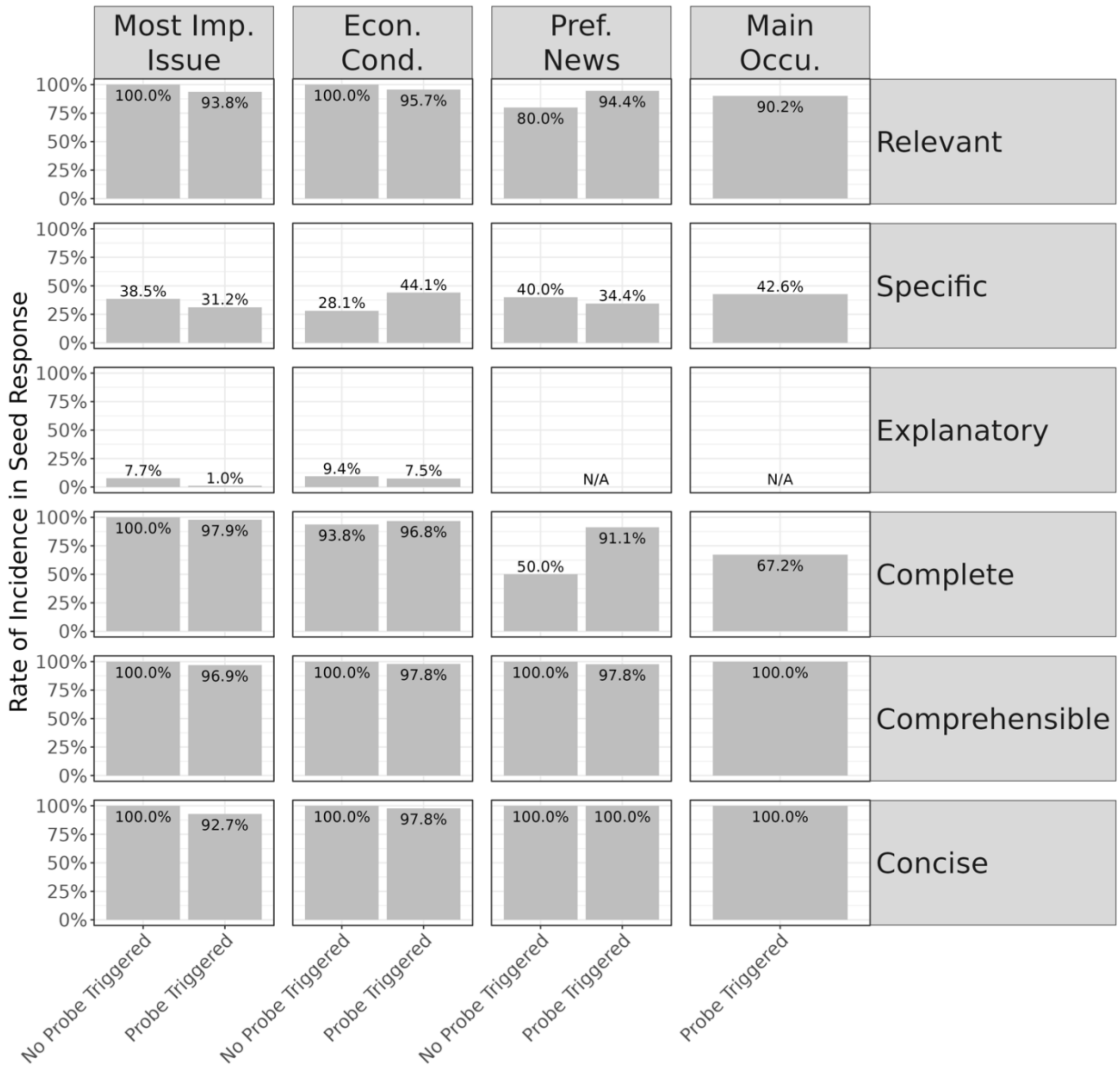

**Figure A14. Effects of Elaboration/Relevance Probing (Treatment 2 Post-Probe vs. Pre-Probe) on Human-Coded Quality Criteria**

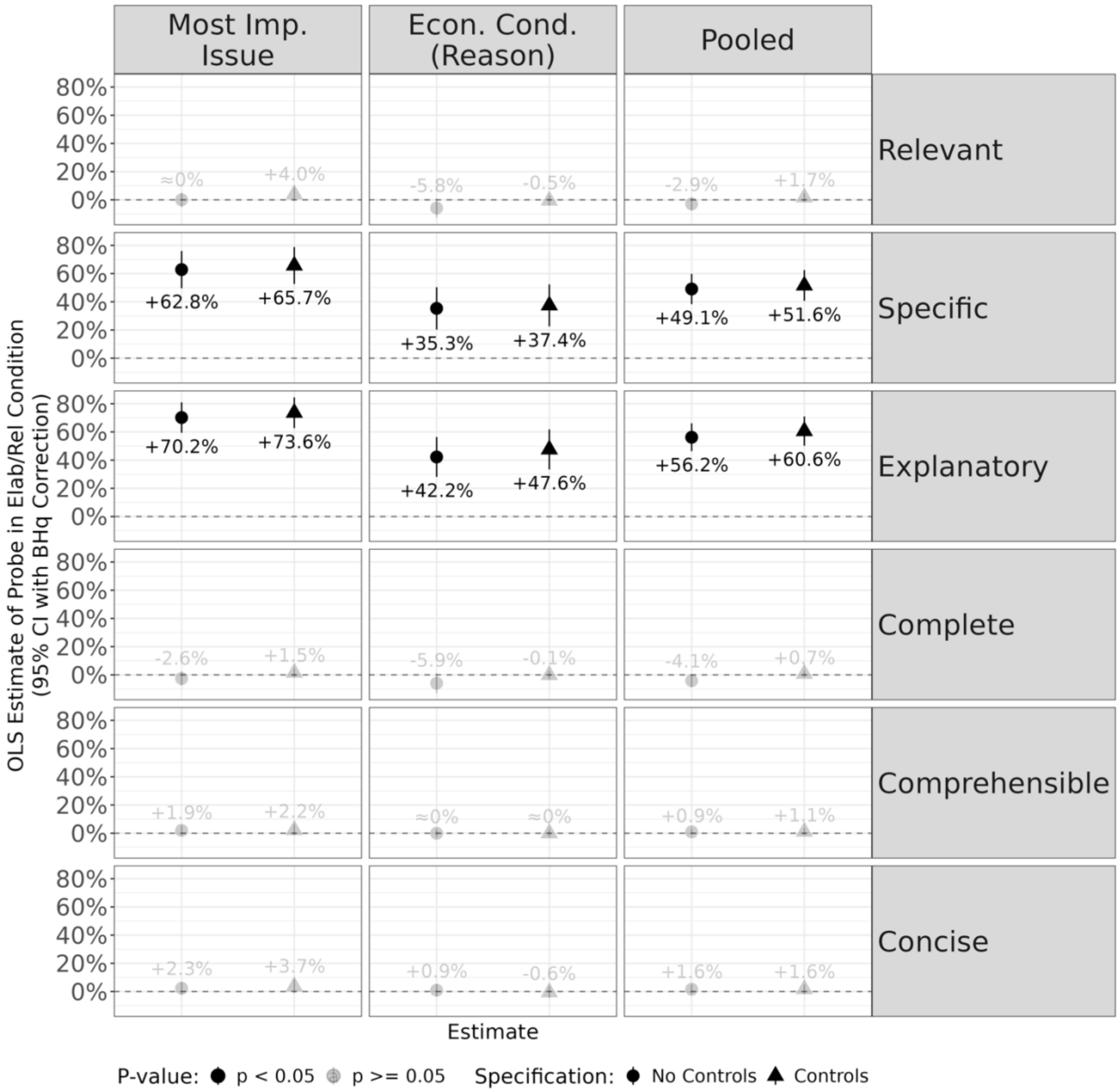

Note: Higher levels correspond to more positive evaluations. Control covariates include work status, gender, education, age, and device type.

## A4. Heterogeneous Treatment Effects of Probing

To capture heterogeneity in response quality effects, we collected demographic information and asked about the survey experience through a series of Likert-scale questions at the end of the survey. For a consistent respondent experience and for validation purposes, we included demographic questions in the survey, even though similar variables were available from the panel provider. Analyses requiring demographic variables (e.g., heterogeneity analyses and multivariate regression models of treatment effects) prioritized the use of panel-provided data when available. In instances where panel data were unavailable, survey-collected variables were used. Our comparison of these sources revealed only minor discrepancies, with a few exceptions. Age was exclusively captured through the survey, introducing the possibility of post-treatment bias if used as a control variable in analyses of treatment effects (Montgomery et al., 2018). We, however, find little to no differences in the regression estimates of treatment effects when including or excluding age as a covariate.

Regression results here follow the same corrections and specifications as in the main text.

**Figure A15. Heterogeneous Effects of Elaboration/Relevance Probing (Treatment 2 vs. Control) on Human-Coded Quality Criteria**

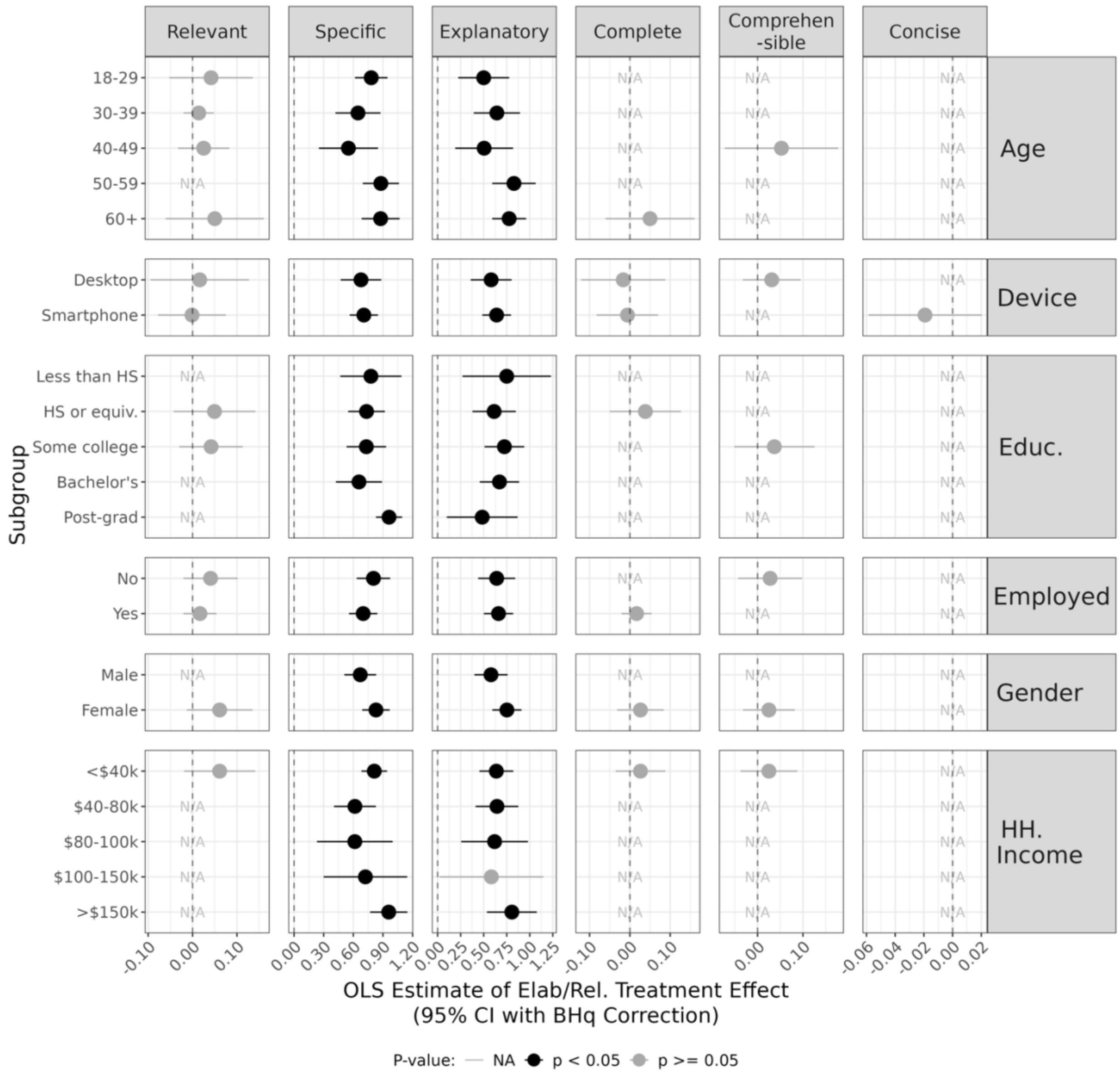

Note: Higher levels correspond to more positive evaluations. Control covariates include work status, gender, education, age, and device type. N/A indicates not a large enough subgroup or enough outcome variation to estimate heterogeneity.

**Figure A16. Heterogeneous Effects of Elaboration/Relevance Probing (Treatment 2) on Self-Reported Respondent Experience**

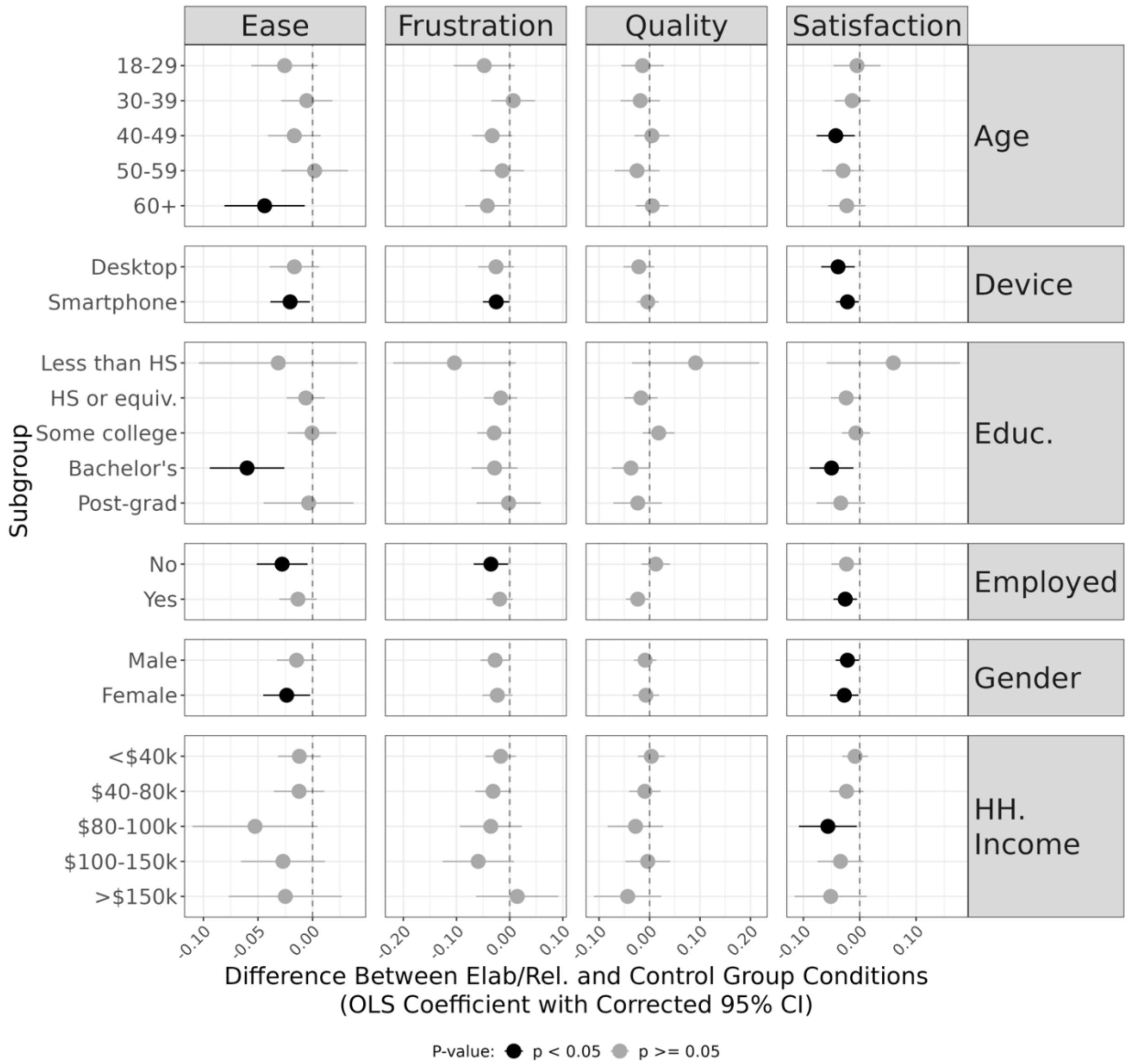

Note: Higher levels correspond to more positive evaluations. Control covariates include work status, gender, education, age, and device type. Outcomes are all normalized to the [0-1] range. Excluded subgroups in 'Other' categories and who answered by tablet due to small sample sizes.